\documentclass[aps,prl,twocolumn,amsmath,amssymb,preprintnumbers,10pt]{revtex4-1}

\usepackage{graphicx}
\usepackage{dcolumn}
\usepackage{bm}
\usepackage{amssymb}
\usepackage{esvect}
\usepackage{xcolor}
\usepackage{braket}
\usepackage{mathtools}
\usepackage{lineno}
\usepackage{color}
\usepackage{soul}
\usepackage{comment}
\usepackage[normalem]{ulem}

\newcommand{\vecr}{{\vec r}}

\newcommand{\veck}{{\vec k}}
\newcommand{\vecb}{{\vec b}}
\newcommand{\vecru}{{\vec {r_1}}}
\newcommand{\vecrd}{{\vec {r_2}}}

\def\nuc#1#2{\relax\ifmmode{}^{#1}{\protect\text{#2}}\else${}^{#1}$#2\fi}

\newcommand{\be}{\begin{eqnarray}}
\newcommand{\ee}{\end{eqnarray}}

\newcommand{\bwt}{\begin{widetext}}
\newcommand{\ewt}{\end{widetext}}

\begin{document}

\title{Isospin dependence in single-nucleon removal cross sections explained through valence-core destruction effects}


\author{M.~G\'omez-Ramos}
\email[]{mgomez40@us.es}

\affiliation{Departamento de FAMN, Universidad de Sevilla, Apartado 1065, 41080 Sevilla, Spain.}

\author{J.~G\'omez-Camacho}

\affiliation{Departamento de FAMN, Universidad de Sevilla,
Apartado 1065, 41080 Sevilla, Spain.}
\affiliation{Centro Nacional de Aceleradores (U. Sevilla, J. Andalucía, CSIC), Tomás Alva Edison, 7, 41092 Sevilla, Spain}

\author{A.M. Moro}

\affiliation{Departamento de FAMN, Universidad de Sevilla,
Apartado 1065, 41080 Sevilla, Spain.}
\affiliation{Instituto Interuniversitario Carlos I de Física Teórica y Computacional (iC1), Apdo. 1065, E-41080 Sevilla, Spain}


\begin{abstract}
The discrepancy between experimental data and theoretical calculations in one-nucleon removal reactions at intermediate energies (quantified by the so-called ``quenching factors'') and its dependence on the isospin asymmetry of the nuclei has been an open problem in nuclear physics for the last fifteen years. 
In this work, we propose an explanation for this long-standing problem, which relies on the inclusion of the process of core destruction due to its interaction with the removed nucleon. To include this effect, we extend the commonly used eikonal formalism via an effective nucleon density, and apply it to a series of nucleon knockout reactions.
The effect of core destruction is found to depend strongly on the binding energy of the removed nucleon, leading to a significant reduction of the cross section for deeply bound nucleons, which reduces the isospin dependence of the ``quenching factors'', making them more consistent with the trends found in transfer and $(p,pN)$ reactions. 
\end{abstract}


\date{\today}%
\maketitle

\label{sec:intro} Single nucleon knockout reactions with  light targets ($^9$Be, $^{12}$C) at intermediate energies have been a key experimental tool to study the structure of unstable nuclei \cite{Orr92,bazin95,simon99,cortina02,cortina04,stroberg15,gade16}. 
These reactions can be described as $P(C+V)+T \rightarrow C +X$, where the projectile $P$ collides with the target $T$ so that the residual nucleus (the core) $C$ is detected, while the valence nucleon $V$ can be detected (diffractive breakup) or is absorbed (stripping). From the momentum distribution of the core, properties of the valence nucleon can be extracted \cite{hufner1981,bertulani92}.
The dynamics of the collision is standardly modelled  within the eikonal approximation \cite{Han03}, which is reasonable for sufficiently high energies ($\sim$80-90 MeV per nucleon). Other nucleon removal reactions such as nucleon transfer \cite{Kay13} and quasifree nucleon removal with proton targets $(p,pN)$ \cite{Jacob:1966} provide complementary information on the properties of the removed nucleons.

A systematic study of the cross section of nucleon knockout reactions in light and medium-mass nuclei showed an intriguing trend \cite{gad08}, where the discrepancy between experimental cross sections and theoretical predictions, quantified by the so-called ``quenching factor'' ($R_s=\sigma_\mathrm{exp}/\sigma_\mathrm{theor}$), shows a marked dependence on the isospin asymmetry of the nucleus, such that for very asymmetric nuclei, the removal of the more abundant nucleons presents a small ``quenching'' ($R_s\sim1$) while the removal of the less abundant ones suffers from a large reduction ($R_s\sim 0.2-0.4$). This tendency has been interpreted as the effect of short-range correlations on the less abundant and more deeply bound nucleons. However, other systematic studies with transfer \cite{Kay13,Fla13,Fla18} and $(p,2p)$ reactions \cite{Ata18,Gom18,hol19} have failed to find this marked dependence on isospin asymmetry, while the addition of new data for heavy-target nucleon-knockout reactions has only reinforced it \cite{Tos14,Tos21}. A recent overview on this topic can be found in \cite{Aum21}. Whether this isospin dependence is a manifestation of short-range correlations not included in standard, small-scale shell-model calculations or an artifact derived from a not yet understood deficiency of the reaction model \cite{Pas20} is a pressing question in nowadays nuclear physics which calls for a careful revision of both the structure and reaction inputs employed in these analyses. 

\begin{figure}[tb]
\begin{center}
 {\centering \resizebox*{0.86\columnwidth}{!}{\includegraphics{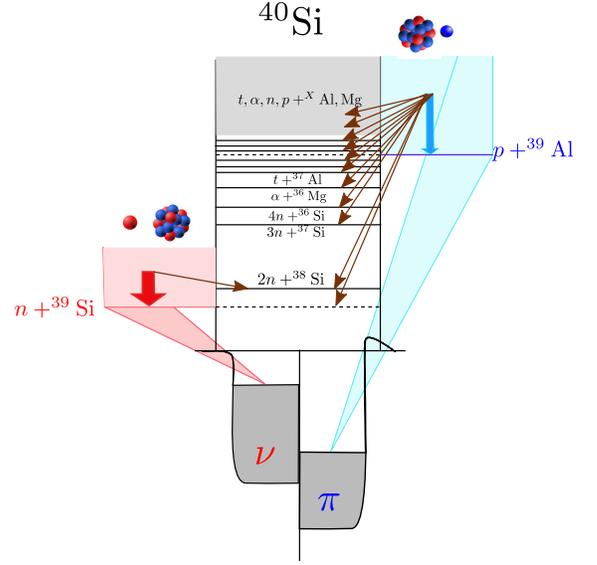}} \par}
\caption{\label{fig:1n_vs_1p}  Schematic illustration of one-proton and one-neutron removal processes in the $^{40}$Si knockout reaction. Note that more channels corresponding to an unbound core are open when a deeply bound nucleon is removed, which leads to larger destruction of the core.}
\end{center}
\end{figure}

Eikonal descriptions assume straight-line trajectories for core and valence nucleon and ignore their mutual final-state interaction. A potentially important effect absent from 
this description of knockout reactions is the destruction of the residual core because of its interaction with the valence nucleon after its removal from the projectile. This core destruction  would naturally lead to a reduction in the knockout cross section, an effect that should be larger when removing more deeply bound nucleons, with stronger interactions with the core, as illustrated in Fig.~\ref{fig:1n_vs_1p}. In fact, intranuclear cascade calculations using the Liege implementation (INCL) \cite{Lou11,Sun16} point to this increased reduction for more deeply-bound nucleons. Moreover, for exclusive breakup reactions, where both valence nucleon and core are detected, the inclusion of this effect in the standard Continuum-Discretized Coupled-Channel (CDCC) formalism \cite{Aus87} has also shown a larger reduction in cross section for the removal of the more deeply-bound species \cite{Gom22}. A recent publication \cite{Heb23} has presented a Green's function description of knockout reactions, but without numerical results.

It is the goal of this work to investigate the effect of these final-state-interactions between the removed nucleon and the residual core on the survival probabilty of the latter in knockout reactions. For that, a novel extension of the eikonal formalism is presented that accounts for such effects and is applied to measured removal reactions for deeply- and weakly-bound nucleons.

 \label{sec:theo}
 
\begin{figure}[tb]
\begin{center}
 {\centering \resizebox*{0.3\columnwidth}{!}{\includegraphics{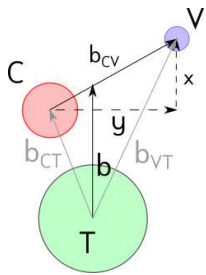}} \par}
\caption{\label{fig:scheme} Scheme of the coordinates used in this work. The beam is perpendicular to the paper.}
\end{center}
\end{figure}

{\it  Theoretical framework}: In the following, we will focus on the 
stripping process. 
The development for diffractive scattering 
 is beyond the scope of this work. 
The stripping channels, although not individually resolved, can be identified by an index $j$ which labels the complex target-nucleon state which, along with the outgoing core, describes the final state. In particular, $j=0$ labels the nucleon-target elastic state where the target remains in its ground state, whereas $j > 0$ correspond to states in which the nucleon excites the target, contributing to stripping. $\vec k$ is the relative momentum between the nucleon and the core. The stripping probability, for some impact parameter $\vecb$, can be written as 
\begin{align}
    P_\mathrm{str}(\vecb) &= \int d^3 \veck \sum_{j \ne 0} |A_j(\vecb, \vec k)|^2 \nonumber \\
     A_j(\vecb, \vec k) & =  \int d^3 \vecr \phi_g(\vecr)^* S^0_{CT}(b_{CT}) S^j_{VT}(b_{VT}) \psi^{(-)}(\vec k, \vec r),
\end{align}
where $\phi_g$ is the bound core-valence state, $S^0_{CT}$ is the core-target elastic S-matrix while $S^j_{VT}$ is the valence-target S-matrix for state $j$ and $\psi^{(-)}(\vec k, \vec r)$ is the final unbound core-valence state. See Fig.~\ref{fig:scheme} for a representation of the impact parameters $b_{CT}$ and $b_{VT}$. A key magnitude in this approach is the nonlocal density:   
\begin{equation}
    \langle \vecrd |\rho_f | \vecru \rangle =  \int d^3 \veck \; \left( \psi^{(-)}(\vec k, \vecrd) \right)^*  \psi^{(-)}(\vec k, \vecru)  .
    \label{closure}
\end{equation}
If the core-nucleon interaction is real (i.e.,  if the nucleon cannot break the core), closure can be used, and $  \langle \vecrd |\rho_f | \vecru \rangle = \delta(\vecru-\vecrd)$. This allows the use of unitarity in the valence-target S-matrix,
\begin{equation}
\sum_{j \ne 0} |S^j_{VT}(b_{VT})|^2 = 1- |S^0_{VT}(b_{VT})|^2, \label{unitarity} \end{equation}
and leads to the standard compact eikonal expression \cite{Han03}
\begin{equation}
P^\mathrm{Ei}_\mathrm{str}(\vecb) = \int d^3 \vec r  \; |\phi_g(\vecr)|^2 |S^0_{CT}(b_{CT})|^2 \left( 1- |S^0_{VT}(b_{VT})|^2 \right),
\end{equation}

 However, in a more realistic situation, where the interaction between valence and core is taken as complex and energy dependent (for example, to describe the excitation or break-up of the core) this is not the case.  This is the key contribution of our work, as compared to standard eikonal approximations. Instead of assuming closure, we will use complex valence-core interactions to get explicitly the continuum wavefunctions at all energies, and then evaluate $ \langle \vecru |\rho_f | \vecrd \rangle$, which would be non local.

In this work, we look for an expression as close as possible to the eikonal derivation. The expression of the proton removal probability requires integration over two radial variables, $\vecru, \vecrd$. The integrand involves the product of two S matrices $S^j_{VT}(b_{VT}(\vecru)) (S^j_{VT}(b_{VT}(\vecrd)))^*$, which are evaluated at different impact parameters, so unitarity (Eq.~(\ref{unitarity})) cannot be applied. This problem with unitarity can be avoided by approximating the two impact parameters in the previous expressions by an average impact parameter defined as $b_{VT}= \sqrt{(b + \alpha x)^2 + (\alpha y)^2}$,
where $\alpha = {A-1 \over A}$, $x= {x_1 + x_2 \over 2}$ and $y = {\sqrt{\dfrac{y_1 ^2 + y_2^2}{2}}}$. $b_{CT}$ requires an equivalent expression. Then we can approximate 
\begin{equation} \sum_{j \ne 0} S^j_{VT}(b_{VT}(\vecru)) (S^j_{VT}(b_{VT}(\vecrd)))^* \simeq 1- |S^0_{VT}(b_{VT}(x,y))|^2, \label{appunitarity} \end{equation}

leading to 
\begin{align}
P_{\rm{str}}(\vecb) & \simeq \int dx dy \; \rho^{(2)\rm{eff}}(x,y) 
\nonumber \\ 
& \times |S^0_{CT}(b_{CT})|^2 \left( 1- |S^0_{VT}(b_{VT})|^2 \right) \\\label{eq:efstr}
\rho^{(2)\rm{eff}}(x,y) &= \int \! d^3 \vecru \int \! d^3 \vecrd \; 
 \langle \vecrd |\rho_f | \vecru \rangle \phi_g^*(\vecrd) \phi_g(\vecru) \nonumber \\
& \times \delta \left(x- {x_1 + x_2 \over 2}\right)  \delta\left(y- \sqrt{\dfrac{y_1^2 + y_2^2}{2}} \right), 
\end{align}
where $\langle \vecrd |\rho_f | \vecru \rangle$ must be computed without applying closure. A more detailed derivation of $\rho^{(2)}_{eff}$ can be found in the Supplementary Material.
Thus, core destruction through interaction with the valence particle can be simply described, in standard eikonal calculations, by using an effective two-dimensional local density $\rho^{(2)\rm{eff}}(x,y)$, which is obtained from the nonlocal final density $\langle \vecrd |\rho_f | \vecru \rangle$ and the nonlocal initial density $\phi_g^*(\vecrd) \phi_g(\vecru) $.  In the usual eikonal approach, this two-dimensional local density is obtained by integrating the ground state density on $z$, giving rise to:
\begin{equation}
\rho^{(2)\rm{Ei}}(x,y) = \int d z |\phi_g(\vecr)|^2. 
\end{equation}
As the impact parameters defining valence and target absorption depend on the coordinates $(x,y)$, the densities required to do the calculations of the stripping probabilities require the two-dimensional densities $\rho^{(2)\rm{Ei}}(x,y)$, $ \rho^{(2)\rm{eff}}(x,y)$. 
One may also compute one-dimensional densities for $x$ (or $y$) as:
\begin{align}
\rho^{(1)\rm{eff}}(x) &= \int \! dy \rho^{(2)\rm{eff}}(x,y) \\
\rho^{(1)\rm{Ei}}(x) &= \int \! dy \rho^{(2)\rm{Ei}}(x,y) .
\end{align}

In the Supplementary Material, an expansion to optimize the calculation of $\rho^{\rm{eff}}$ is presented. A fundamental difference between this method and standard eikonal calculations (e.g. \cite{Sau04}) lies in the consideration that valence particle and core keep interacting after the absorption of the former by the target, while standard calculations neglect this interaction. We believe that this interaction is still important for the dynamics of the reaction even after the valence particle has been absorbed (as it has not disappeared, rather it has become deeply correlated with the internal degrees of freedom of the target), which is consistent with the spirit and results from INCL calculations \cite{Lou11,Sun16}.

 \label{sec:results}
 
\begin{figure}[tb]
\begin{center}
 {\centering \resizebox*{\columnwidth}{!}{\includegraphics{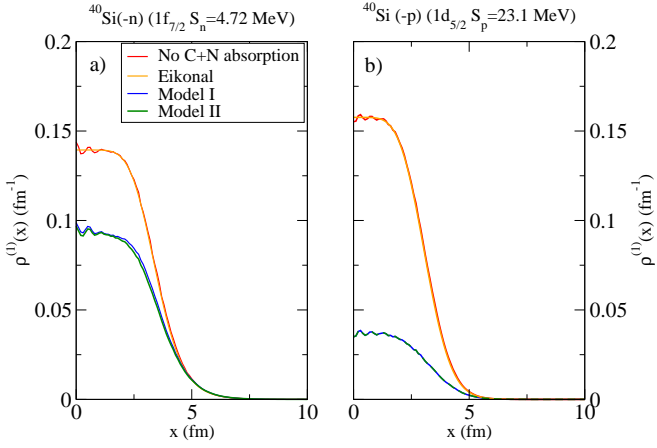}} \par}
\caption{\label{fig:dens40si}One-dimensional effective density for the valence neutron (left) and proton (right) in the $1f_{7/2}$ and $1d_{5/2}$ orbitals, respectively, for $^{40}$Si. The red line corresponds to the effective density without core destruction, and the orange one to the eikonal calculation. The blue and green curves correspond to the calculation with core destruction, for model I and model II, respectively. The potential required no modification for the valence proton, so both curves coincide in the right panel (see text).}
\end{center}
\end{figure}

{\it Results}: We apply this formalism to a selection of the knockout reactions presented in \cite{Tos21}, for removal of neutron and proton from a neutron-rich nucleus ($^{40}$Si), a proton-rich nucleus ($^{24}$Si) and an isospin symmetric one ($^{12}$C). These nuclei were selected because both proton and neutron removal were measured, only a few single-particle configurations of the removed nucleon had to be considered (except for neutron removal from $^{40}$Si) and because the ingredients for the original calculations were accessible in the literature. In order to restrict the integration in $\vec{k}$ in the evaluation of $\langle \vecrd |\rho_f | \vecru \rangle$, we included a weighting factor $e^{-k^4a^4}$ with $a=0.15$ fm$^{-1}$ and expanded 
$\psi^{(-)}(\vec k, \vec{r})$ in multipoles up to $l_\mathrm{max}=29$ (see Supplementary Material). For the single-particle wavefunction we used the same geometry \cite{gad08,bro02,Strthesis} used in the results presented in \cite{gad08,Tos14,Tos21} for $^{24}$Si, $^{12}$C and $^{40}$Si respectively. To build the continuum wavefunctions, an optical potential is required. For consistency and to focus on core destruction, for the evaluation of this potential, we have considered the imaginary part of the global energy-dependent dispersive potential by Morillon \textit{et al.} \cite{Mor07} for all considered nuclei. This potential reproduces reasonably the reaction cross sections from the ENDF database \cite{Bro18} between $p-^{11}$B and $n-^{11}$C for energies $>20$~MeV. As is general for optical potentials, the computed reaction cross section (related to the imaginary part of the potential) includes the formation of compound nucleus, which may decay into the original valence-core channel, not resulting in the destruction of the core. Therefore, for a proper description of the destruction of the core, the potential must be modified to eliminate this process from the reaction cross section. In order to evaluate the importance of this ``elastic-compound-nucleus'' contribution, we have performed compound-nucleus calculations to obtain the fraction of the cross section that results in actual destruction of the core for the different systems in a range of relevant energies. Then, for the different energies, we have rescaled the reaction cross sections obtained with the Morillon potential by this factor and modified the depth of the imaginary surface term of the potential to reproduce this core-destruction cross section (when required, the imaginary volume term was removed) (see supplementary material). Given the significant dispersion in compound nucleus results \cite{Bla18}, we present the results using two widely-used compound-nucleus codes: PACE \cite{Tar08,Gav80}, which will be referred to as Model I, and GEMINI \cite{Cha10,Man10}, which will be referred to as Model II. The effects of the neglect of elastic compound nucleus are presented in the Supplementary Material. We note that for the deeply bound nucleons many open channels exist even at zero relative energy (as illustrated in Fig.~\ref{fig:1n_vs_1p}) so the elastic channel was not significantly populated in the compound-nucleus calculations and no potential modification was required. The same occurred for all nuclei at valence-core energies $>30$ MeV.  The computed effective density is presented as a function of $x$ in Fig.~\ref{fig:dens40si}, where the left panel corresponds to the valence neutron in $^{40}$Si in the $1f_{7/2}$ orbital (bound by 4.72 MeV) and the right panel to the valence proton in the $1d_{5/2}$ orbital (bound by 23.1 MeV). To validate the density calculation, the red line corresponds to calculations where $\psi^{(-)}(\vec k, \vec{r})$ were taken as plane waves, which should coincide with the density for the eikonal calculation corresponding to the orange line.
 For both cases, the plane-wave and eikonal calculations agree very well, except for small oscillations in the interior, which can be related to the  cutoff in $k$ and $l$. When comparing the plane-wave calculation to that with core destruction (the blue line corresponds to model I and the green line to model II), core destruction is shown to produce a significant reduction in the density for both cases, particularly in the interior. This reduction is larger for the more bound case (less abundant species), as expected due to the abundance of open channels (see Fig.~\ref{fig:1n_vs_1p}).

\begin{figure}[tb]
\begin{center}
 {\centering \resizebox*{\columnwidth}{!}{\includegraphics{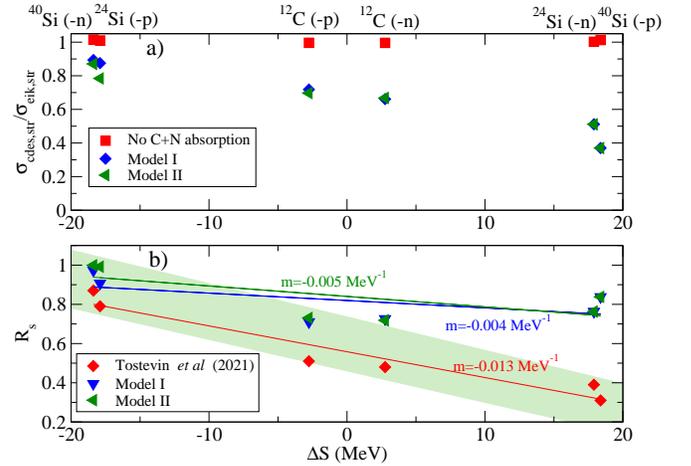}} \par}
\caption{\label{fig:gade} a) Ratio between computed stripping cross sections and standard eikonal calculations \cite{Han03} as a function of the difference in proton-neutron separation energy $\Delta S$. Calculations are shown without the $N+C$ potential (red squares) and with core destruction modelled with model I (blue diamonds) and with model II (green triangles). b) Standard (red diamonds) and modified ``quenching factors'' $R_s$ as a function of $\Delta S$, considering for the $N+C$ system model I (blue triangles) and model II (green triangles). (See text).}
\end{center}
\end{figure}

To evaluate the effect of this reduction on stripping cross sections, the latter have been computed using the effective density  from Eq.~(\ref{eq:efstr}). The values for $S^0_{CT}(b_{CT})$ and $S^0_{VT}(b_{VT})$ have been taken from the original references. The top panel of Fig.~\ref{fig:gade} shows ratios between the computed stripping cross sections and those from the standard eikonal model \cite{Han03} as a function of the difference between the separation energy of the removed species and its isospin pair $\Delta S=S_{n(p)}-S_{p(n)}$ \cite{gad08}, with $S_{n(p)}$ taken from \cite{nudat}. For all cases except $^{40}$Si$(-n)$, only one single-particle configuration was dominant in the cross section. For $^{40}$Si$(-n)$, the $1f_{7/2},2p_{3/2},1d_{5/2}$ and $2s_{1/2}$ configurations were considered and weighted by their spectroscopic factors from the SDPF-U interaction \cite{Strthesis}, which accounts for 95\% of the cross section. Red squares correspond to the effective density computed without core destruction, with a difference to the standard calculation of at most $ 1.5\%$. Blue diamonds and green triangles correspond to the calculations using models I and II, respectively. The results show larger reduction for removal of the more deeply-bound nucleon (to the right of the graph), $\sim 0.5$ and smaller reduction in the weakly-bound case, $\sim 0.9$. The reduction in cross section is smaller than the one in the norm of the density, as seen in Fig.~\ref{fig:dens40si}, due to the peripherality of the reaction, since in the nuclear surface the reduction due to core destruction is smaller than in the interior.

The bottom panel of Fig.~\ref{fig:gade} shows the effect of this reduction on the ``quenching factors'' $R_s$. Red diamonds correspond to the original values from \cite{Tos21}. Since we have only studied the effect of core destruction in stripping, to compare to experimental data, which also include diffractive scattering, we will assume the same reduction for diffractive scattering. Since stripping is the main contributor to the cross section \cite{gad08,bro02,Strthesis}, we consider this approximation to be sufficient for the purposes of this work. Therefore, the values of $R_s$ with core destruction are computed through:
\begin{equation}
    R_s^\mathrm{cdes}={\sigma_\mathrm{exp}\over\sigma_\mathrm{cdes}} ={\sigma_\mathrm{exp} \over \sigma_\mathrm{eik}} {\sigma_\mathrm{eik} \over \sigma_\mathrm{cdes}} \simeq  {\sigma_\mathrm{exp} \over \sigma_\mathrm{eik}} {\sigma_\mathrm{eik,str} \over \sigma_\mathrm{cdes,str}}=\dfrac{R_s^\mathrm{orig}}{\dfrac{\sigma_\mathrm{cdes,str}}{\sigma_\mathrm{eik,str}}},\label{eq:Rs_mod}
\end{equation}
 (where $\sigma_\mathrm{exp},\sigma_\mathrm{eik},\sigma_\mathrm{cdes}$ are the experimental, standard eikonal and with-core-destruction knockout cross sections and $\sigma_\mathrm{eik,str}, \sigma_\mathrm{cdes,str}$ the ones for stripping) and are presented in the bottom panel of Fig.~\ref{fig:gade} as blue and green triangles, corresponding to calculations using models I and II, respectively. These modified ``quenching'' factors present a significantly smaller dependence on $\Delta S$, with a slope of $-0.004 \mathrm{MeV}^{-1}$ for model I and $-0.005 \mathrm{MeV}^{-1}$ for model II, which is less than half the original value: $-0.013 \mathrm{MeV}^{-1}$.
Therefore, these results indicate that a large part of the dependence of the ``quenching factors'' on $\Delta S$ can be related to the destruction of the core through its interaction with the removed particle, an effect that can be included in standard eikonal calculations using the effective density from Eq.~(\ref{eq:efstr}). Including core destruction significantly reduces the dependence on isospin asymmetry, making the trend for nucleon-knockout reactions consistent with that in transfer and $(p,pN)$ reactions. It is remarkable that the tendency with core destruction agrees quite well with the reduction in spectroscopic factors found in coupled-cluster calculations for oxygen isotopes \cite{Jen11}, which would suggest that the remaining dependence on $\Delta S$ could be described by many-body correlations. These results show that the low-energy interaction between removed particle and core is fundamental to properly interpret the measurements from nucleon-knockout experiments. Therefore, better information on this interaction, obtained from theoretical calculations starting from first principles or from measurement of nucleon-core reaction cross sections (particularly for exotic species with larger $|\Delta S|$), is essential to extract significant spectroscopic information from nucleon knockout experiments. 

{\it Summary and outlook}: In this work, we have investigated the effect of core destruction due to its final-state interaction with the removed nucleon in nucleon stripping reactions. The inclusion of this effect significantly flattens the dependence of the ``quenching factors'' on isospin asymmetry, making this dependence consistent with that found in transfer and $(p,pN)$ reactions.  Therefore, core destruction appears as 
one of the key contributors to answer the open question on this dependence. Experimental measurements that detect the products of core destruction could be used as validation of these results. A precedent exists for these measurements: in \cite{Sun16} experimental results were compared to INCL calculations for nucleon removal from $^{14}$O. As well, confirmation of these results would require more accurate optical potentials between valence nucleon and core, which could be extracted via ab-initio methods \cite{Idi19}. Experimental measurements to extract the core-nucleon reaction cross section would also be useful to reduce the uncertainties in the potentials required for these calculations. Some improvements in the formalism are also desirable, such as an extension to diffractive scattering or a more sophisticated description of the reaction going beyond the eikonal approximation, with proper energy and momentum conservation, such as the Ichimura-Austern-Vincent (IAV) formalism \cite{IAV85,Jin15,Jin15b,Pot15,Car16}, which could be extended to include valence-core destruction. In addition, the inclusion of the real part of the valence-core interaction (and its bound states), which has been neglected in this work, should be considered. Further work on the latter points is currently underway.

\begin{acknowledgements}
The authors thank A.~Di Pietro for her help in the calculations with PACE4 and GEMINI. M.G.-R.\, J.G.-C.\ and A.M.M.\ acknowledge financial support by MCIN/AEI/10.13039/501100011033 under I+D+i project No.\ PID2020-114687GB-I00 and under grant IJC2020-043878-I (also funded by ``European Union NextGenerationEU/PRTR''), by the Consejer\'{\i}a de Econom\'{\i}a, Conocimiento, Empresas y Universidad, Junta de Andaluc\'{\i}a (Spain) and ``ERDF-A Way of Making Europe'' under PAIDI 2020 project No.\ P20\_01247, and by the European Social Fund and Junta de Andalucía (PAIDI 2020) under grant number DOC-01006.
\end{acknowledgements}
 


\bibliography{inclusive.bib}

\begin{thebibliography}{45}%
\makeatletter
\providecommand \@ifxundefined [1]{%
 \@ifx{#1\undefined}
}%
\providecommand \@ifnum [1]{%
 \ifnum #1\expandafter \@firstoftwo
 \else \expandafter \@secondoftwo
 \fi
}%
\providecommand \@ifx [1]{%
 \ifx #1\expandafter \@firstoftwo
 \else \expandafter \@secondoftwo
 \fi
}%
\providecommand \natexlab [1]{#1}%
\providecommand \enquote  [1]{``#1''}%
\providecommand \bibnamefont  [1]{#1}%
\providecommand \bibfnamefont [1]{#1}%
\providecommand \citenamefont [1]{#1}%
\providecommand \href@noop [0]{\@secondoftwo}%
\providecommand \href [0]{\begingroup \@sanitize@url \@href}%
\providecommand \@href[1]{\@@startlink{#1}\@@href}%
\providecommand \@@href[1]{\endgroup#1\@@endlink}%
\providecommand \@sanitize@url [0]{\catcode `\\12\catcode `\$12\catcode
  `\&12\catcode `\#12\catcode `\^12\catcode `\_12\catcode `\%12\relax}%
\providecommand \@@startlink[1]{}%
\providecommand \@@endlink[0]{}%
\providecommand \url  [0]{\begingroup\@sanitize@url \@url }%
\providecommand \@url [1]{\endgroup\@href {#1}{\urlprefix }}%
\providecommand \urlprefix  [0]{URL }%
\providecommand \Eprint [0]{\href }%
\providecommand \doibase [0]{https://doi.org/}%
\providecommand \selectlanguage [0]{\@gobble}%
\providecommand \bibinfo  [0]{\@secondoftwo}%
\providecommand \bibfield  [0]{\@secondoftwo}%
\providecommand \translation [1]{[#1]}%
\providecommand \BibitemOpen [0]{}%
\providecommand \bibitemStop [0]{}%
\providecommand \bibitemNoStop [0]{.\EOS\space}%
\providecommand \EOS [0]{\spacefactor3000\relax}%
\providecommand \BibitemShut  [1]{\csname bibitem#1\endcsname}%
\let\auto@bib@innerbib\@empty
\bibitem [{\citenamefont {Orr}\ \emph {et~al.}(1992)\citenamefont {Orr},
  \citenamefont {Anantaraman}, \citenamefont {Austin}, \citenamefont
  {Bertulani}, \citenamefont {Hanold}, \citenamefont {Kelley}, \citenamefont
  {Morrissey}, \citenamefont {Sherrill}, \citenamefont {Souliotis},
  \citenamefont {Thoennessen}, \citenamefont {Winfield},\ and\ \citenamefont
  {Winger}}]{Orr92}%
  \BibitemOpen
  \bibfield  {author} {\bibinfo {author} {\bibfnamefont {N.~A.}\ \bibnamefont
  {Orr}}, \bibinfo {author} {\bibfnamefont {N.}~\bibnamefont {Anantaraman}},
  \bibinfo {author} {\bibfnamefont {S.~M.}\ \bibnamefont {Austin}}, \bibinfo
  {author} {\bibfnamefont {C.~A.}\ \bibnamefont {Bertulani}}, \bibinfo {author}
  {\bibfnamefont {K.}~\bibnamefont {Hanold}}, \bibinfo {author} {\bibfnamefont
  {J.~H.}\ \bibnamefont {Kelley}}, \bibinfo {author} {\bibfnamefont {D.~J.}\
  \bibnamefont {Morrissey}}, \bibinfo {author} {\bibfnamefont {B.~M.}\
  \bibnamefont {Sherrill}}, \bibinfo {author} {\bibfnamefont {G.~A.}\
  \bibnamefont {Souliotis}}, \bibinfo {author} {\bibfnamefont {M.}~\bibnamefont
  {Thoennessen}}, \bibinfo {author} {\bibfnamefont {J.~S.}\ \bibnamefont
  {Winfield}},\ and\ \bibinfo {author} {\bibfnamefont {J.~A.}\ \bibnamefont
  {Winger}},\ }\href {https://doi.org/10.1103/PhysRevLett.69.2050} {\bibfield
  {journal} {\bibinfo  {journal} {Phys. Rev. Lett.}\ }\textbf {\bibinfo
  {volume} {69}},\ \bibinfo {pages} {2050} (\bibinfo {year}
  {1992})}\BibitemShut {NoStop}%
\bibitem [{\citenamefont {Bazin}\ \emph {et~al.}(1995)\citenamefont {Bazin},
  \citenamefont {Brown}, \citenamefont {Brown}, \citenamefont {Fauerbach},
  \citenamefont {Hellstr\"om}, \citenamefont {Hirzebruch}, \citenamefont
  {Kelley}, \citenamefont {Kryger}, \citenamefont {Morrissey}, \citenamefont
  {Pfaff}, \citenamefont {Powell}, \citenamefont {Sherrill},\ and\
  \citenamefont {Thoennessen}}]{bazin95}%
  \BibitemOpen
  \bibfield  {author} {\bibinfo {author} {\bibfnamefont {D.}~\bibnamefont
  {Bazin}}, \bibinfo {author} {\bibfnamefont {B.~A.}\ \bibnamefont {Brown}},
  \bibinfo {author} {\bibfnamefont {J.}~\bibnamefont {Brown}}, \bibinfo
  {author} {\bibfnamefont {M.}~\bibnamefont {Fauerbach}}, \bibinfo {author}
  {\bibfnamefont {M.}~\bibnamefont {Hellstr\"om}}, \bibinfo {author}
  {\bibfnamefont {S.~E.}\ \bibnamefont {Hirzebruch}}, \bibinfo {author}
  {\bibfnamefont {J.~H.}\ \bibnamefont {Kelley}}, \bibinfo {author}
  {\bibfnamefont {R.~A.}\ \bibnamefont {Kryger}}, \bibinfo {author}
  {\bibfnamefont {D.~J.}\ \bibnamefont {Morrissey}}, \bibinfo {author}
  {\bibfnamefont {R.}~\bibnamefont {Pfaff}}, \bibinfo {author} {\bibfnamefont
  {C.~F.}\ \bibnamefont {Powell}}, \bibinfo {author} {\bibfnamefont {B.~M.}\
  \bibnamefont {Sherrill}},\ and\ \bibinfo {author} {\bibfnamefont
  {M.}~\bibnamefont {Thoennessen}},\ }\href
  {https://link.aps.org/doi/10.1103/PhysRevLett.74.3569} {\bibfield  {journal}
  {\bibinfo  {journal} {Phys. Rev. Lett.}\ }\textbf {\bibinfo {volume} {74}},\
  \bibinfo {pages} {3569} (\bibinfo {year} {1995})}\BibitemShut {NoStop}%
\bibitem [{\citenamefont {Simon}\ \emph {et~al.}(1999)\citenamefont {Simon},
  \citenamefont {Aleksandrov}, \citenamefont {Aumann}, \citenamefont
  {Axelsson}, \citenamefont {Baumann}, \citenamefont {Borge}, \citenamefont
  {Chulkov}, \citenamefont {Collatz}, \citenamefont {Cub}, \citenamefont
  {Dostal}, \citenamefont {Eberlein}, \citenamefont {Elze}, \citenamefont
  {Emling}, \citenamefont {Geissel}, \citenamefont {Gr\"unschloss},
  \citenamefont {Hellstr\"om}, \citenamefont {Holeczek}, \citenamefont
  {Holzmann}, \citenamefont {Jonson}, \citenamefont {Kratz}, \citenamefont
  {Kraus}, \citenamefont {Kulessa}, \citenamefont {Leifels}, \citenamefont
  {Leistenschneider}, \citenamefont {Leth}, \citenamefont {Mukha},
  \citenamefont {M\"unzenberg}, \citenamefont {Nickel}, \citenamefont
  {Nilsson}, \citenamefont {Nyman}, \citenamefont {Petersen}, \citenamefont
  {Pf\"utzner}, \citenamefont {Richter}, \citenamefont {Riisager},
  \citenamefont {Scheidenberger}, \citenamefont {Schrieder}, \citenamefont
  {Schwab}, \citenamefont {Smedberg}, \citenamefont {Stroth}, \citenamefont
  {Surowiec}, \citenamefont {Tengblad},\ and\ \citenamefont
  {Zhukov}}]{simon99}%
  \BibitemOpen
  \bibfield  {author} {\bibinfo {author} {\bibfnamefont {H.}~\bibnamefont
  {Simon}}, \bibinfo {author} {\bibfnamefont {D.}~\bibnamefont {Aleksandrov}},
  \bibinfo {author} {\bibfnamefont {T.}~\bibnamefont {Aumann}}, \bibinfo
  {author} {\bibfnamefont {L.}~\bibnamefont {Axelsson}}, \bibinfo {author}
  {\bibfnamefont {T.}~\bibnamefont {Baumann}}, \bibinfo {author} {\bibfnamefont
  {M.~J.~G.}\ \bibnamefont {Borge}}, \bibinfo {author} {\bibfnamefont {L.~V.}\
  \bibnamefont {Chulkov}}, \bibinfo {author} {\bibfnamefont {R.}~\bibnamefont
  {Collatz}}, \bibinfo {author} {\bibfnamefont {J.}~\bibnamefont {Cub}},
  \bibinfo {author} {\bibfnamefont {W.}~\bibnamefont {Dostal}}, \bibinfo
  {author} {\bibfnamefont {B.}~\bibnamefont {Eberlein}}, \bibinfo {author}
  {\bibfnamefont {T.~W.}\ \bibnamefont {Elze}}, \bibinfo {author}
  {\bibfnamefont {H.}~\bibnamefont {Emling}}, \bibinfo {author} {\bibfnamefont
  {H.}~\bibnamefont {Geissel}}, \bibinfo {author} {\bibfnamefont
  {A.}~\bibnamefont {Gr\"unschloss}}, \bibinfo {author} {\bibfnamefont
  {M.}~\bibnamefont {Hellstr\"om}}, \bibinfo {author} {\bibfnamefont
  {J.}~\bibnamefont {Holeczek}}, \bibinfo {author} {\bibfnamefont
  {R.}~\bibnamefont {Holzmann}}, \bibinfo {author} {\bibfnamefont
  {B.}~\bibnamefont {Jonson}}, \bibinfo {author} {\bibfnamefont {J.~V.}\
  \bibnamefont {Kratz}}, \bibinfo {author} {\bibfnamefont {G.}~\bibnamefont
  {Kraus}}, \bibinfo {author} {\bibfnamefont {R.}~\bibnamefont {Kulessa}},
  \bibinfo {author} {\bibfnamefont {Y.}~\bibnamefont {Leifels}}, \bibinfo
  {author} {\bibfnamefont {A.}~\bibnamefont {Leistenschneider}}, \bibinfo
  {author} {\bibfnamefont {T.}~\bibnamefont {Leth}}, \bibinfo {author}
  {\bibfnamefont {I.}~\bibnamefont {Mukha}}, \bibinfo {author} {\bibfnamefont
  {G.}~\bibnamefont {M\"unzenberg}}, \bibinfo {author} {\bibfnamefont
  {F.}~\bibnamefont {Nickel}}, \bibinfo {author} {\bibfnamefont
  {T.}~\bibnamefont {Nilsson}}, \bibinfo {author} {\bibfnamefont
  {G.}~\bibnamefont {Nyman}}, \bibinfo {author} {\bibfnamefont
  {B.}~\bibnamefont {Petersen}}, \bibinfo {author} {\bibfnamefont
  {M.}~\bibnamefont {Pf\"utzner}}, \bibinfo {author} {\bibfnamefont
  {A.}~\bibnamefont {Richter}}, \bibinfo {author} {\bibfnamefont
  {K.}~\bibnamefont {Riisager}}, \bibinfo {author} {\bibfnamefont
  {C.}~\bibnamefont {Scheidenberger}}, \bibinfo {author} {\bibfnamefont
  {G.}~\bibnamefont {Schrieder}}, \bibinfo {author} {\bibfnamefont
  {W.}~\bibnamefont {Schwab}}, \bibinfo {author} {\bibfnamefont {M.~H.}\
  \bibnamefont {Smedberg}}, \bibinfo {author} {\bibfnamefont {J.}~\bibnamefont
  {Stroth}}, \bibinfo {author} {\bibfnamefont {A.}~\bibnamefont {Surowiec}},
  \bibinfo {author} {\bibfnamefont {O.}~\bibnamefont {Tengblad}},\ and\
  \bibinfo {author} {\bibfnamefont {M.~V.}\ \bibnamefont {Zhukov}},\ }\href
  {https://doi.org/10.1103/PhysRevLett.83.496} {\bibfield  {journal} {\bibinfo
  {journal} {Phys. Rev. Lett.}\ }\textbf {\bibinfo {volume} {83}},\ \bibinfo
  {pages} {496} (\bibinfo {year} {1999})}\BibitemShut {NoStop}%
\bibitem [{\citenamefont {Cortina-Gil}\ \emph {et~al.}(2002)\citenamefont
  {Cortina-Gil}, \citenamefont {Markenroth}, \citenamefont {Attallah},
  \citenamefont {Baumann}, \citenamefont {Benlliure}, \citenamefont {Borge},
  \citenamefont {Chulkov}, \citenamefont {Pramanik}, \citenamefont
  {Fernandez-Vazquez}, \citenamefont {Forssen}, \citenamefont {Fraile},
  \citenamefont {Geissel}, \citenamefont {Gerl}, \citenamefont {Hammache},
  \citenamefont {Itahashi}, \citenamefont {Janik}, \citenamefont {Jonson},
  \citenamefont {Karlsson}, \citenamefont {Lenske}, \citenamefont {Mandal},
  \citenamefont {Meister}, \citenamefont {Mocko}, \citenamefont
  {M{\"u}nzenberg}, \citenamefont {Ohtsubo}, \citenamefont {Ozawa},
  \citenamefont {Parfenova}, \citenamefont {Pribora}, \citenamefont {Riisager},
  \citenamefont {Scheit}, \citenamefont {Schneider}, \citenamefont {Schmidt},
  \citenamefont {Schrieder}, \citenamefont {Simon}, \citenamefont {Sitar},
  \citenamefont {Stolz}, \citenamefont {Strmen}, \citenamefont {S{\"u}mmerer},
  \citenamefont {Szarka}, \citenamefont {Wan}, \citenamefont {Weick},\ and\
  \citenamefont {Zhukov}}]{cortina02}%
  \BibitemOpen
  \bibfield  {author} {\bibinfo {author} {\bibfnamefont {D.}~\bibnamefont
  {Cortina-Gil}}, \bibinfo {author} {\bibfnamefont {K.}~\bibnamefont
  {Markenroth}}, \bibinfo {author} {\bibfnamefont {F.}~\bibnamefont
  {Attallah}}, \bibinfo {author} {\bibfnamefont {T.}~\bibnamefont {Baumann}},
  \bibinfo {author} {\bibfnamefont {J.}~\bibnamefont {Benlliure}}, \bibinfo
  {author} {\bibfnamefont {M.}~\bibnamefont {Borge}}, \bibinfo {author}
  {\bibfnamefont {L.}~\bibnamefont {Chulkov}}, \bibinfo {author} {\bibfnamefont
  {U.~D.}\ \bibnamefont {Pramanik}}, \bibinfo {author} {\bibfnamefont
  {J.}~\bibnamefont {Fernandez-Vazquez}}, \bibinfo {author} {\bibfnamefont
  {C.}~\bibnamefont {Forssen}}, \bibinfo {author} {\bibfnamefont
  {L.}~\bibnamefont {Fraile}}, \bibinfo {author} {\bibfnamefont
  {H.}~\bibnamefont {Geissel}}, \bibinfo {author} {\bibfnamefont
  {J.}~\bibnamefont {Gerl}}, \bibinfo {author} {\bibfnamefont {F.}~\bibnamefont
  {Hammache}}, \bibinfo {author} {\bibfnamefont {K.}~\bibnamefont {Itahashi}},
  \bibinfo {author} {\bibfnamefont {R.}~\bibnamefont {Janik}}, \bibinfo
  {author} {\bibfnamefont {B.}~\bibnamefont {Jonson}}, \bibinfo {author}
  {\bibfnamefont {S.}~\bibnamefont {Karlsson}}, \bibinfo {author}
  {\bibfnamefont {H.}~\bibnamefont {Lenske}}, \bibinfo {author} {\bibfnamefont
  {S.}~\bibnamefont {Mandal}}, \bibinfo {author} {\bibfnamefont
  {M.}~\bibnamefont {Meister}}, \bibinfo {author} {\bibfnamefont
  {X.}~\bibnamefont {Mocko}}, \bibinfo {author} {\bibfnamefont
  {G.}~\bibnamefont {M{\"u}nzenberg}}, \bibinfo {author} {\bibfnamefont
  {T.}~\bibnamefont {Ohtsubo}}, \bibinfo {author} {\bibfnamefont
  {A.}~\bibnamefont {Ozawa}}, \bibinfo {author} {\bibfnamefont
  {Y.}~\bibnamefont {Parfenova}}, \bibinfo {author} {\bibfnamefont
  {V.}~\bibnamefont {Pribora}}, \bibinfo {author} {\bibfnamefont
  {K.}~\bibnamefont {Riisager}}, \bibinfo {author} {\bibfnamefont
  {H.}~\bibnamefont {Scheit}}, \bibinfo {author} {\bibfnamefont
  {R.}~\bibnamefont {Schneider}}, \bibinfo {author} {\bibfnamefont
  {K.}~\bibnamefont {Schmidt}}, \bibinfo {author} {\bibfnamefont
  {G.}~\bibnamefont {Schrieder}}, \bibinfo {author} {\bibfnamefont
  {H.}~\bibnamefont {Simon}}, \bibinfo {author} {\bibfnamefont
  {B.}~\bibnamefont {Sitar}}, \bibinfo {author} {\bibfnamefont
  {A.}~\bibnamefont {Stolz}}, \bibinfo {author} {\bibfnamefont
  {P.}~\bibnamefont {Strmen}}, \bibinfo {author} {\bibfnamefont
  {K.}~\bibnamefont {S{\"u}mmerer}}, \bibinfo {author} {\bibfnamefont
  {I.}~\bibnamefont {Szarka}}, \bibinfo {author} {\bibfnamefont
  {S.}~\bibnamefont {Wan}}, \bibinfo {author} {\bibfnamefont {H.}~\bibnamefont
  {Weick}},\ and\ \bibinfo {author} {\bibfnamefont {M.}~\bibnamefont
  {Zhukov}},\ }\href
  {https://doi.org/https://doi.org/10.1016/S0370-2693(02)01245-5} {\bibfield
  {journal} {\bibinfo  {journal} {Physics Letters B}\ }\textbf {\bibinfo
  {volume} {529}},\ \bibinfo {pages} {36 } (\bibinfo {year}
  {2002})}\BibitemShut {NoStop}%
\bibitem [{\citenamefont {Cortina-Gil}\ \emph {et~al.}(2004)\citenamefont
  {Cortina-Gil}, \citenamefont {Fernandez-Vazquez}, \citenamefont {Aumann},
  \citenamefont {Baumann}, \citenamefont {Benlliure}, \citenamefont {Borge},
  \citenamefont {Chulkov}, \citenamefont {Datta~Pramanik}, \citenamefont
  {Forss\'en}, \citenamefont {Fraile}, \citenamefont {Geissel}, \citenamefont
  {Gerl}, \citenamefont {Hammache}, \citenamefont {Itahashi}, \citenamefont
  {Janik}, \citenamefont {Jonson}, \citenamefont {Mandal}, \citenamefont
  {Markenroth}, \citenamefont {Meister}, \citenamefont {Mocko}, \citenamefont
  {M\"unzenberg}, \citenamefont {Ohtsubo}, \citenamefont {Ozawa}, \citenamefont
  {Prezado}, \citenamefont {Pribora}, \citenamefont {Riisager}, \citenamefont
  {Scheit}, \citenamefont {Schneider}, \citenamefont {Schrieder}, \citenamefont
  {Simon}, \citenamefont {Sitar}, \citenamefont {Stolz}, \citenamefont
  {Strmen}, \citenamefont {S\"ummerer}, \citenamefont {Szarka},\ and\
  \citenamefont {Weick}}]{cortina04}%
  \BibitemOpen
  \bibfield  {author} {\bibinfo {author} {\bibfnamefont {D.}~\bibnamefont
  {Cortina-Gil}}, \bibinfo {author} {\bibfnamefont {J.}~\bibnamefont
  {Fernandez-Vazquez}}, \bibinfo {author} {\bibfnamefont {T.}~\bibnamefont
  {Aumann}}, \bibinfo {author} {\bibfnamefont {T.}~\bibnamefont {Baumann}},
  \bibinfo {author} {\bibfnamefont {J.}~\bibnamefont {Benlliure}}, \bibinfo
  {author} {\bibfnamefont {M.~J.~G.}\ \bibnamefont {Borge}}, \bibinfo {author}
  {\bibfnamefont {L.~V.}\ \bibnamefont {Chulkov}}, \bibinfo {author}
  {\bibfnamefont {U.}~\bibnamefont {Datta~Pramanik}}, \bibinfo {author}
  {\bibfnamefont {C.}~\bibnamefont {Forss\'en}}, \bibinfo {author}
  {\bibfnamefont {L.~M.}\ \bibnamefont {Fraile}}, \bibinfo {author}
  {\bibfnamefont {H.}~\bibnamefont {Geissel}}, \bibinfo {author} {\bibfnamefont
  {J.}~\bibnamefont {Gerl}}, \bibinfo {author} {\bibfnamefont {F.}~\bibnamefont
  {Hammache}}, \bibinfo {author} {\bibfnamefont {K.}~\bibnamefont {Itahashi}},
  \bibinfo {author} {\bibfnamefont {R.}~\bibnamefont {Janik}}, \bibinfo
  {author} {\bibfnamefont {B.}~\bibnamefont {Jonson}}, \bibinfo {author}
  {\bibfnamefont {S.}~\bibnamefont {Mandal}}, \bibinfo {author} {\bibfnamefont
  {K.}~\bibnamefont {Markenroth}}, \bibinfo {author} {\bibfnamefont
  {M.}~\bibnamefont {Meister}}, \bibinfo {author} {\bibfnamefont
  {M.}~\bibnamefont {Mocko}}, \bibinfo {author} {\bibfnamefont
  {G.}~\bibnamefont {M\"unzenberg}}, \bibinfo {author} {\bibfnamefont
  {T.}~\bibnamefont {Ohtsubo}}, \bibinfo {author} {\bibfnamefont
  {A.}~\bibnamefont {Ozawa}}, \bibinfo {author} {\bibfnamefont
  {Y.}~\bibnamefont {Prezado}}, \bibinfo {author} {\bibfnamefont
  {V.}~\bibnamefont {Pribora}}, \bibinfo {author} {\bibfnamefont
  {K.}~\bibnamefont {Riisager}}, \bibinfo {author} {\bibfnamefont
  {H.}~\bibnamefont {Scheit}}, \bibinfo {author} {\bibfnamefont
  {R.}~\bibnamefont {Schneider}}, \bibinfo {author} {\bibfnamefont
  {G.}~\bibnamefont {Schrieder}}, \bibinfo {author} {\bibfnamefont
  {H.}~\bibnamefont {Simon}}, \bibinfo {author} {\bibfnamefont
  {B.}~\bibnamefont {Sitar}}, \bibinfo {author} {\bibfnamefont
  {A.}~\bibnamefont {Stolz}}, \bibinfo {author} {\bibfnamefont
  {P.}~\bibnamefont {Strmen}}, \bibinfo {author} {\bibfnamefont
  {K.}~\bibnamefont {S\"ummerer}}, \bibinfo {author} {\bibfnamefont
  {I.}~\bibnamefont {Szarka}},\ and\ \bibinfo {author} {\bibfnamefont
  {H.}~\bibnamefont {Weick}},\ }\href
  {https://doi.org/10.1103/PhysRevLett.93.062501} {\bibfield  {journal}
  {\bibinfo  {journal} {Phys. Rev. Lett.}\ }\textbf {\bibinfo {volume} {93}},\
  \bibinfo {pages} {062501} (\bibinfo {year} {2004})}\BibitemShut {NoStop}%
\bibitem [{\citenamefont {Stroberg}\ \emph {et~al.}(2015)\citenamefont
  {Stroberg}, \citenamefont {Gade}, \citenamefont {Tostevin}, \citenamefont
  {Bader}, \citenamefont {Baugher}, \citenamefont {Bazin}, \citenamefont
  {Berryman}, \citenamefont {Brown}, \citenamefont {Campbell}, \citenamefont
  {Kemper}, \citenamefont {Langer}, \citenamefont {Lunderberg}, \citenamefont
  {Lemasson}, \citenamefont {Noji}, \citenamefont {Otsuka}, \citenamefont
  {Recchia}, \citenamefont {Walz}, \citenamefont {Weisshaar},\ and\
  \citenamefont {Williams}}]{stroberg15}%
  \BibitemOpen
  \bibfield  {author} {\bibinfo {author} {\bibfnamefont {S.~R.}\ \bibnamefont
  {Stroberg}}, \bibinfo {author} {\bibfnamefont {A.}~\bibnamefont {Gade}},
  \bibinfo {author} {\bibfnamefont {J.~A.}\ \bibnamefont {Tostevin}}, \bibinfo
  {author} {\bibfnamefont {V.~M.}\ \bibnamefont {Bader}}, \bibinfo {author}
  {\bibfnamefont {T.}~\bibnamefont {Baugher}}, \bibinfo {author} {\bibfnamefont
  {D.}~\bibnamefont {Bazin}}, \bibinfo {author} {\bibfnamefont {J.~S.}\
  \bibnamefont {Berryman}}, \bibinfo {author} {\bibfnamefont {B.~A.}\
  \bibnamefont {Brown}}, \bibinfo {author} {\bibfnamefont {C.~M.}\ \bibnamefont
  {Campbell}}, \bibinfo {author} {\bibfnamefont {K.~W.}\ \bibnamefont
  {Kemper}}, \bibinfo {author} {\bibfnamefont {C.}~\bibnamefont {Langer}},
  \bibinfo {author} {\bibfnamefont {E.}~\bibnamefont {Lunderberg}}, \bibinfo
  {author} {\bibfnamefont {A.}~\bibnamefont {Lemasson}}, \bibinfo {author}
  {\bibfnamefont {S.}~\bibnamefont {Noji}}, \bibinfo {author} {\bibfnamefont
  {T.}~\bibnamefont {Otsuka}}, \bibinfo {author} {\bibfnamefont
  {F.}~\bibnamefont {Recchia}}, \bibinfo {author} {\bibfnamefont
  {C.}~\bibnamefont {Walz}}, \bibinfo {author} {\bibfnamefont {D.}~\bibnamefont
  {Weisshaar}},\ and\ \bibinfo {author} {\bibfnamefont {S.}~\bibnamefont
  {Williams}},\ }\href {https://doi.org/10.1103/PhysRevC.91.041302} {\bibfield
  {journal} {\bibinfo  {journal} {Phys. Rev. C}\ }\textbf {\bibinfo {volume}
  {91}},\ \bibinfo {pages} {041302} (\bibinfo {year} {2015})}\BibitemShut
  {NoStop}%
\bibitem [{\citenamefont {Gade}\ \emph {et~al.}(2016)\citenamefont {Gade},
  \citenamefont {Tostevin}, \citenamefont {Bader}, \citenamefont {Baugher},
  \citenamefont {Bazin}, \citenamefont {Berryman}, \citenamefont {Brown},
  \citenamefont {Diget}, \citenamefont {Glasmacher}, \citenamefont {Hartley},
  \citenamefont {Lunderberg}, \citenamefont {Stroberg}, \citenamefont
  {Recchia}, \citenamefont {Ratkiewicz}, \citenamefont {Weisshaar},\ and\
  \citenamefont {Wimmer}}]{gade16}%
  \BibitemOpen
  \bibfield  {author} {\bibinfo {author} {\bibfnamefont {A.}~\bibnamefont
  {Gade}}, \bibinfo {author} {\bibfnamefont {J.~A.}\ \bibnamefont {Tostevin}},
  \bibinfo {author} {\bibfnamefont {V.}~\bibnamefont {Bader}}, \bibinfo
  {author} {\bibfnamefont {T.}~\bibnamefont {Baugher}}, \bibinfo {author}
  {\bibfnamefont {D.}~\bibnamefont {Bazin}}, \bibinfo {author} {\bibfnamefont
  {J.~S.}\ \bibnamefont {Berryman}}, \bibinfo {author} {\bibfnamefont {B.~A.}\
  \bibnamefont {Brown}}, \bibinfo {author} {\bibfnamefont {C.~A.}\ \bibnamefont
  {Diget}}, \bibinfo {author} {\bibfnamefont {T.}~\bibnamefont {Glasmacher}},
  \bibinfo {author} {\bibfnamefont {D.~J.}\ \bibnamefont {Hartley}}, \bibinfo
  {author} {\bibfnamefont {E.}~\bibnamefont {Lunderberg}}, \bibinfo {author}
  {\bibfnamefont {S.~R.}\ \bibnamefont {Stroberg}}, \bibinfo {author}
  {\bibfnamefont {F.}~\bibnamefont {Recchia}}, \bibinfo {author} {\bibfnamefont
  {A.}~\bibnamefont {Ratkiewicz}}, \bibinfo {author} {\bibfnamefont
  {D.}~\bibnamefont {Weisshaar}},\ and\ \bibinfo {author} {\bibfnamefont
  {K.}~\bibnamefont {Wimmer}},\ }\href
  {https://doi.org/10.1103/PhysRevC.93.054315} {\bibfield  {journal} {\bibinfo
  {journal} {Phys. Rev. C}\ }\textbf {\bibinfo {volume} {93}},\ \bibinfo
  {pages} {054315} (\bibinfo {year} {2016})}\BibitemShut {NoStop}%
\bibitem [{\citenamefont {H\"ufner}\ and\ \citenamefont
  {Nemes}(1981)}]{hufner1981}%
  \BibitemOpen
  \bibfield  {author} {\bibinfo {author} {\bibfnamefont {J.}~\bibnamefont
  {H\"ufner}}\ and\ \bibinfo {author} {\bibfnamefont {M.~C.}\ \bibnamefont
  {Nemes}},\ }\bibfield  {journal} {\bibinfo  {journal} {Phys. Rev. C}\
  }\textbf {\bibinfo {volume} {23}},\ \href
  {https://doi.org/10.1103/PhysRevC.23.2538} {10.1103/PhysRevC.23.2538}
  (\bibinfo {year} {1981})\BibitemShut {NoStop}%
\bibitem [{\citenamefont {Bertulani}\ and\ \citenamefont
  {McVoy}(1992)}]{bertulani92}%
  \BibitemOpen
  \bibfield  {author} {\bibinfo {author} {\bibfnamefont {C.~A.}\ \bibnamefont
  {Bertulani}}\ and\ \bibinfo {author} {\bibfnamefont {K.~W.}\ \bibnamefont
  {McVoy}},\ }\href {https://doi.org/10.1103/PhysRevC.46.2638} {\bibfield
  {journal} {\bibinfo  {journal} {Phys. Rev. C}\ }\textbf {\bibinfo {volume}
  {46}},\ \bibinfo {pages} {2638} (\bibinfo {year} {1992})}\BibitemShut
  {NoStop}%
\bibitem [{\citenamefont {Hansen}\ and\ \citenamefont
  {Tostevin}(2003)}]{Han03}%
  \BibitemOpen
  \bibfield  {author} {\bibinfo {author} {\bibfnamefont {P.}~\bibnamefont
  {Hansen}}\ and\ \bibinfo {author} {\bibfnamefont {J.}~\bibnamefont
  {Tostevin}},\ }\href {https://doi.org/10.1146/annurev.nucl.53.041002.110406}
  {\bibfield  {journal} {\bibinfo  {journal} {Annu. Rev. of Nucl. and Part.
  Sci}\ }\textbf {\bibinfo {volume} {53}},\ \bibinfo {pages} {219} (\bibinfo
  {year} {2003})}\BibitemShut {NoStop}%
\bibitem [{\citenamefont {Kay}\ \emph {et~al.}(2013)\citenamefont {Kay},
  \citenamefont {Schiffer},\ and\ \citenamefont {Freeman}}]{Kay13}%
  \BibitemOpen
  \bibfield  {author} {\bibinfo {author} {\bibfnamefont {B.~P.}\ \bibnamefont
  {Kay}}, \bibinfo {author} {\bibfnamefont {J.~P.}\ \bibnamefont {Schiffer}},\
  and\ \bibinfo {author} {\bibfnamefont {S.~J.}\ \bibnamefont {Freeman}},\
  }\href {https://doi.org/10.1103/PhysRevLett.111.042502} {\bibfield  {journal}
  {\bibinfo  {journal} {Phys. Rev. Lett.}\ }\textbf {\bibinfo {volume} {111}},\
  \bibinfo {pages} {042502} (\bibinfo {year} {2013})}\BibitemShut {NoStop}%
\bibitem [{\citenamefont {Jacob}\ and\ \citenamefont
  {Maris}(1966)}]{Jacob:1966}%
  \BibitemOpen
  \bibfield  {author} {\bibinfo {author} {\bibfnamefont {G.}~\bibnamefont
  {Jacob}}\ and\ \bibinfo {author} {\bibfnamefont {T.~A.~J.}\ \bibnamefont
  {Maris}},\ }\href {https://doi.org/10.1103/RevModPhys.38.121} {\bibfield
  {journal} {\bibinfo  {journal} {Rev. Mod. Phys.}\ }\textbf {\bibinfo {volume}
  {38}},\ \bibinfo {pages} {121} (\bibinfo {year} {1966})}\BibitemShut
  {NoStop}%
\bibitem [{\citenamefont {Gade}\ \emph {et~al.}(2008)\citenamefont {Gade},
  \citenamefont {Adrich}, \citenamefont {Bazin}, \citenamefont {Bowen},
  \citenamefont {Brown}, \citenamefont {Campbell}, \citenamefont {Cook},
  \citenamefont {Glasmacher}, \citenamefont {Hansen}, \citenamefont {Hosier},
  \citenamefont {McDaniel}, \citenamefont {McGlinchery}, \citenamefont
  {Obertelli}, \citenamefont {Siwek}, \citenamefont {Riley}, \citenamefont
  {Tostevin},\ and\ \citenamefont {Weisshaar}}]{gad08}%
  \BibitemOpen
  \bibfield  {author} {\bibinfo {author} {\bibfnamefont {A.}~\bibnamefont
  {Gade}}, \bibinfo {author} {\bibfnamefont {P.}~\bibnamefont {Adrich}},
  \bibinfo {author} {\bibfnamefont {D.}~\bibnamefont {Bazin}}, \bibinfo
  {author} {\bibfnamefont {M.~D.}\ \bibnamefont {Bowen}}, \bibinfo {author}
  {\bibfnamefont {B.~A.}\ \bibnamefont {Brown}}, \bibinfo {author}
  {\bibfnamefont {C.~M.}\ \bibnamefont {Campbell}}, \bibinfo {author}
  {\bibfnamefont {J.~M.}\ \bibnamefont {Cook}}, \bibinfo {author}
  {\bibfnamefont {T.}~\bibnamefont {Glasmacher}}, \bibinfo {author}
  {\bibfnamefont {P.~G.}\ \bibnamefont {Hansen}}, \bibinfo {author}
  {\bibfnamefont {K.}~\bibnamefont {Hosier}}, \bibinfo {author} {\bibfnamefont
  {S.}~\bibnamefont {McDaniel}}, \bibinfo {author} {\bibfnamefont
  {D.}~\bibnamefont {McGlinchery}}, \bibinfo {author} {\bibfnamefont
  {A.}~\bibnamefont {Obertelli}}, \bibinfo {author} {\bibfnamefont
  {K.}~\bibnamefont {Siwek}}, \bibinfo {author} {\bibfnamefont {L.~A.}\
  \bibnamefont {Riley}}, \bibinfo {author} {\bibfnamefont {J.~A.}\ \bibnamefont
  {Tostevin}},\ and\ \bibinfo {author} {\bibfnamefont {D.}~\bibnamefont
  {Weisshaar}},\ }\href {https://doi.org/10.1103/PhysRevC.77.044306} {\bibfield
   {journal} {\bibinfo  {journal} {Phys. Rev. C}\ }\textbf {\bibinfo {volume}
  {77}},\ \bibinfo {pages} {044306} (\bibinfo {year} {2008})}\BibitemShut
  {NoStop}%
\bibitem [{\citenamefont {Flavigny}\ \emph {et~al.}(2013)\citenamefont
  {Flavigny} \emph {et~al.}}]{Fla13}%
  \BibitemOpen
  \bibfield  {author} {\bibinfo {author} {\bibfnamefont {F.}~\bibnamefont
  {Flavigny}} \emph {et~al.},\ }\href
  {https://doi.org/10.1103/PhysRevLett.110.122503} {\bibfield  {journal}
  {\bibinfo  {journal} {Phys. Rev. Lett.}\ }\textbf {\bibinfo {volume} {110}},\
  \bibinfo {pages} {122503} (\bibinfo {year} {2013})}\BibitemShut {NoStop}%
\bibitem [{\citenamefont {Flavigny}\ \emph {et~al.}(2018)\citenamefont
  {Flavigny}, \citenamefont {Keeley}, \citenamefont {Gillibert},\ and\
  \citenamefont {Obertelli}}]{Fla18}%
  \BibitemOpen
  \bibfield  {author} {\bibinfo {author} {\bibfnamefont {F.}~\bibnamefont
  {Flavigny}}, \bibinfo {author} {\bibfnamefont {N.}~\bibnamefont {Keeley}},
  \bibinfo {author} {\bibfnamefont {A.}~\bibnamefont {Gillibert}},\ and\
  \bibinfo {author} {\bibfnamefont {A.}~\bibnamefont {Obertelli}},\ }\href
  {https://doi.org/10.1103/PhysRevC.97.034601} {\bibfield  {journal} {\bibinfo
  {journal} {Phys. Rev. C}\ }\textbf {\bibinfo {volume} {97}},\ \bibinfo
  {pages} {034601} (\bibinfo {year} {2018})}\BibitemShut {NoStop}%
\bibitem [{\citenamefont {Atar}\ \emph {et~al.}(2018)\citenamefont {Atar} \emph
  {et~al.}}]{Ata18}%
  \BibitemOpen
  \bibfield  {author} {\bibinfo {author} {\bibfnamefont {L.}~\bibnamefont
  {Atar}} \emph {et~al.} (\bibinfo {collaboration} {R$^{3}$B Collaboration}),\
  }\href {https://doi.org/10.1103/PhysRevLett.120.052501} {\bibfield  {journal}
  {\bibinfo  {journal} {Phys. Rev. Lett.}\ }\textbf {\bibinfo {volume} {120}},\
  \bibinfo {pages} {052501} (\bibinfo {year} {2018})}\BibitemShut {NoStop}%
\bibitem [{\citenamefont {Gómez-Ramos}\ and\ \citenamefont
  {Moro}(2018)}]{Gom18}%
  \BibitemOpen
  \bibfield  {author} {\bibinfo {author} {\bibfnamefont {M.}~\bibnamefont
  {Gómez-Ramos}}\ and\ \bibinfo {author} {\bibfnamefont {A.}~\bibnamefont
  {Moro}},\ }\href {https://doi.org/10.1016/j.physletb.2018.08.058} {\bibfield
  {journal} {\bibinfo  {journal} {Phys. Lett. B}\ }\textbf {\bibinfo {volume}
  {785}},\ \bibinfo {pages} {511 } (\bibinfo {year} {2018})}\BibitemShut
  {NoStop}%
\bibitem [{\citenamefont {Holl}\ \emph {et~al.}(2019)\citenamefont {Holl} \emph
  {et~al.}}]{hol19}%
  \BibitemOpen
  \bibfield  {author} {\bibinfo {author} {\bibfnamefont {M.}~\bibnamefont
  {Holl}} \emph {et~al.},\ }\href
  {https://doi.org/https://doi.org/10.1016/j.physletb.2019.06.069} {\bibfield
  {journal} {\bibinfo  {journal} {Physics Letters B}\ }\textbf {\bibinfo
  {volume} {795}},\ \bibinfo {pages} {682} (\bibinfo {year}
  {2019})}\BibitemShut {NoStop}%
\bibitem [{\citenamefont {Tostevin}\ and\ \citenamefont {Gade}(2014)}]{Tos14}%
  \BibitemOpen
  \bibfield  {author} {\bibinfo {author} {\bibfnamefont {J.~A.}\ \bibnamefont
  {Tostevin}}\ and\ \bibinfo {author} {\bibfnamefont {A.}~\bibnamefont
  {Gade}},\ }\href {https://doi.org/10.1103/PhysRevC.90.057602} {\bibfield
  {journal} {\bibinfo  {journal} {Phys. Rev. C}\ }\textbf {\bibinfo {volume}
  {90}},\ \bibinfo {pages} {057602} (\bibinfo {year} {2014})}\BibitemShut
  {NoStop}%
\bibitem [{\citenamefont {Tostevin}\ and\ \citenamefont {Gade}(2021)}]{Tos21}%
  \BibitemOpen
  \bibfield  {author} {\bibinfo {author} {\bibfnamefont {J.~A.}\ \bibnamefont
  {Tostevin}}\ and\ \bibinfo {author} {\bibfnamefont {A.}~\bibnamefont
  {Gade}},\ }\href {https://doi.org/10.1103/PhysRevC.103.054610} {\bibfield
  {journal} {\bibinfo  {journal} {Phys. Rev. C}\ }\textbf {\bibinfo {volume}
  {103}},\ \bibinfo {pages} {054610} (\bibinfo {year} {2021})}\BibitemShut
  {NoStop}%
\bibitem [{\citenamefont {Aumann}\ \emph {et~al.}(2021)\citenamefont {Aumann},
  \citenamefont {Barbieri}, \citenamefont {Bazin}, \citenamefont {Bertulani},
  \citenamefont {Bonaccorso}, \citenamefont {Dickhoff}, \citenamefont {Gade},
  \citenamefont {Gómez-Ramos}, \citenamefont {Kay}, \citenamefont {Moro},
  \citenamefont {Nakamura}, \citenamefont {Obertelli}, \citenamefont {Ogata},
  \citenamefont {Paschalis},\ and\ \citenamefont {Uesaka}}]{Aum21}%
  \BibitemOpen
  \bibfield  {author} {\bibinfo {author} {\bibfnamefont {T.}~\bibnamefont
  {Aumann}}, \bibinfo {author} {\bibfnamefont {C.}~\bibnamefont {Barbieri}},
  \bibinfo {author} {\bibfnamefont {D.}~\bibnamefont {Bazin}}, \bibinfo
  {author} {\bibfnamefont {C.}~\bibnamefont {Bertulani}}, \bibinfo {author}
  {\bibfnamefont {A.}~\bibnamefont {Bonaccorso}}, \bibinfo {author}
  {\bibfnamefont {W.}~\bibnamefont {Dickhoff}}, \bibinfo {author}
  {\bibfnamefont {A.}~\bibnamefont {Gade}}, \bibinfo {author} {\bibfnamefont
  {M.}~\bibnamefont {Gómez-Ramos}}, \bibinfo {author} {\bibfnamefont
  {B.}~\bibnamefont {Kay}}, \bibinfo {author} {\bibfnamefont {A.}~\bibnamefont
  {Moro}}, \bibinfo {author} {\bibfnamefont {T.}~\bibnamefont {Nakamura}},
  \bibinfo {author} {\bibfnamefont {A.}~\bibnamefont {Obertelli}}, \bibinfo
  {author} {\bibfnamefont {K.}~\bibnamefont {Ogata}}, \bibinfo {author}
  {\bibfnamefont {S.}~\bibnamefont {Paschalis}},\ and\ \bibinfo {author}
  {\bibfnamefont {T.}~\bibnamefont {Uesaka}},\ }\href
  {https://doi.org/doi.org/10.1016/j.ppnp.2021.103847} {\bibfield  {journal}
  {\bibinfo  {journal} {Progress in Particle and Nuclear Physics}\ }\textbf
  {\bibinfo {volume} {118}},\ \bibinfo {pages} {103847} (\bibinfo {year}
  {2021})}\BibitemShut {NoStop}%
\bibitem [{\citenamefont {Paschalis}\ \emph {et~al.}(2020)\citenamefont
  {Paschalis}, \citenamefont {Petri}, \citenamefont {Macchiavelli},
  \citenamefont {Hen},\ and\ \citenamefont {Piasetzky}}]{Pas20}%
  \BibitemOpen
  \bibfield  {author} {\bibinfo {author} {\bibfnamefont {S.}~\bibnamefont
  {Paschalis}}, \bibinfo {author} {\bibfnamefont {M.}~\bibnamefont {Petri}},
  \bibinfo {author} {\bibfnamefont {A.}~\bibnamefont {Macchiavelli}}, \bibinfo
  {author} {\bibfnamefont {O.}~\bibnamefont {Hen}},\ and\ \bibinfo {author}
  {\bibfnamefont {E.}~\bibnamefont {Piasetzky}},\ }\href
  {https://doi.org/https://doi.org/10.1016/j.physletb.2019.135110} {\bibfield
  {journal} {\bibinfo  {journal} {Physics Letters B}\ }\textbf {\bibinfo
  {volume} {800}},\ \bibinfo {pages} {135110} (\bibinfo {year}
  {2020})}\BibitemShut {NoStop}%
\bibitem [{\citenamefont {Louchart}\ \emph {et~al.}(2011)\citenamefont
  {Louchart}, \citenamefont {Obertelli}, \citenamefont {Boudard},\ and\
  \citenamefont {Flavigny}}]{Lou11}%
  \BibitemOpen
  \bibfield  {author} {\bibinfo {author} {\bibfnamefont {C.}~\bibnamefont
  {Louchart}}, \bibinfo {author} {\bibfnamefont {A.}~\bibnamefont {Obertelli}},
  \bibinfo {author} {\bibfnamefont {A.}~\bibnamefont {Boudard}},\ and\ \bibinfo
  {author} {\bibfnamefont {F.}~\bibnamefont {Flavigny}},\ }\href
  {https://doi.org/10.1103/PhysRevC.83.011601} {\bibfield  {journal} {\bibinfo
  {journal} {Phys. Rev. C}\ }\textbf {\bibinfo {volume} {83}},\ \bibinfo
  {pages} {011601} (\bibinfo {year} {2011})}\BibitemShut {NoStop}%
\bibitem [{\citenamefont {Sun}\ \emph {et~al.}(2016)\citenamefont {Sun},
  \citenamefont {Lee}, \citenamefont {Ye}, \citenamefont {Obertelli},
  \citenamefont {Li}, \citenamefont {Aoi}, \citenamefont {Ong}, \citenamefont
  {Ayyad}, \citenamefont {Bertulani}, \citenamefont {Chen}, \citenamefont
  {Corsi}, \citenamefont {Cappuzzello}, \citenamefont {Cavallaro},
  \citenamefont {Furono}, \citenamefont {Ge}, \citenamefont {Hashimoto},
  \citenamefont {Ideguchi}, \citenamefont {Kawabata}, \citenamefont {Lou},
  \citenamefont {Li}, \citenamefont {Lorusso}, \citenamefont {Lu},
  \citenamefont {Liu}, \citenamefont {Nishimura}, \citenamefont {Suzuki},
  \citenamefont {Tanaka}, \citenamefont {Tanaka}, \citenamefont {Tran},
  \citenamefont {Tsang}, \citenamefont {Wu}, \citenamefont {Xu},\ and\
  \citenamefont {Yamamoto}}]{Sun16}%
  \BibitemOpen
  \bibfield  {author} {\bibinfo {author} {\bibfnamefont {Y.~L.}\ \bibnamefont
  {Sun}}, \bibinfo {author} {\bibfnamefont {J.}~\bibnamefont {Lee}}, \bibinfo
  {author} {\bibfnamefont {Y.~L.}\ \bibnamefont {Ye}}, \bibinfo {author}
  {\bibfnamefont {A.}~\bibnamefont {Obertelli}}, \bibinfo {author}
  {\bibfnamefont {Z.~H.}\ \bibnamefont {Li}}, \bibinfo {author} {\bibfnamefont
  {N.}~\bibnamefont {Aoi}}, \bibinfo {author} {\bibfnamefont {H.~J.}\
  \bibnamefont {Ong}}, \bibinfo {author} {\bibfnamefont {Y.}~\bibnamefont
  {Ayyad}}, \bibinfo {author} {\bibfnamefont {C.~A.}\ \bibnamefont
  {Bertulani}}, \bibinfo {author} {\bibfnamefont {J.}~\bibnamefont {Chen}},
  \bibinfo {author} {\bibfnamefont {A.}~\bibnamefont {Corsi}}, \bibinfo
  {author} {\bibfnamefont {F.}~\bibnamefont {Cappuzzello}}, \bibinfo {author}
  {\bibfnamefont {M.}~\bibnamefont {Cavallaro}}, \bibinfo {author}
  {\bibfnamefont {T.}~\bibnamefont {Furono}}, \bibinfo {author} {\bibfnamefont
  {Y.~C.}\ \bibnamefont {Ge}}, \bibinfo {author} {\bibfnamefont
  {T.}~\bibnamefont {Hashimoto}}, \bibinfo {author} {\bibfnamefont
  {E.}~\bibnamefont {Ideguchi}}, \bibinfo {author} {\bibfnamefont
  {T.}~\bibnamefont {Kawabata}}, \bibinfo {author} {\bibfnamefont {J.~L.}\
  \bibnamefont {Lou}}, \bibinfo {author} {\bibfnamefont {Q.~T.}\ \bibnamefont
  {Li}}, \bibinfo {author} {\bibfnamefont {G.}~\bibnamefont {Lorusso}},
  \bibinfo {author} {\bibfnamefont {F.}~\bibnamefont {Lu}}, \bibinfo {author}
  {\bibfnamefont {H.~N.}\ \bibnamefont {Liu}}, \bibinfo {author} {\bibfnamefont
  {S.}~\bibnamefont {Nishimura}}, \bibinfo {author} {\bibfnamefont
  {H.}~\bibnamefont {Suzuki}}, \bibinfo {author} {\bibfnamefont
  {J.}~\bibnamefont {Tanaka}}, \bibinfo {author} {\bibfnamefont
  {M.}~\bibnamefont {Tanaka}}, \bibinfo {author} {\bibfnamefont {D.~T.}\
  \bibnamefont {Tran}}, \bibinfo {author} {\bibfnamefont {M.~B.}\ \bibnamefont
  {Tsang}}, \bibinfo {author} {\bibfnamefont {J.}~\bibnamefont {Wu}}, \bibinfo
  {author} {\bibfnamefont {Z.~Y.}\ \bibnamefont {Xu}},\ and\ \bibinfo {author}
  {\bibfnamefont {T.}~\bibnamefont {Yamamoto}},\ }\href
  {https://doi.org/10.1103/PhysRevC.93.044607} {\bibfield  {journal} {\bibinfo
  {journal} {Phys. Rev. C}\ }\textbf {\bibinfo {volume} {93}},\ \bibinfo
  {pages} {044607} (\bibinfo {year} {2016})}\BibitemShut {NoStop}%
\bibitem [{\citenamefont {Austern}\ \emph {et~al.}(1987)\citenamefont
  {Austern}, \citenamefont {Iseri}, \citenamefont {Kamimura}, \citenamefont
  {Kawai}, \citenamefont {Rawitscher},\ and\ \citenamefont {Yahiro}}]{Aus87}%
  \BibitemOpen
  \bibfield  {author} {\bibinfo {author} {\bibfnamefont {N.}~\bibnamefont
  {Austern}}, \bibinfo {author} {\bibfnamefont {Y.}~\bibnamefont {Iseri}},
  \bibinfo {author} {\bibfnamefont {M.}~\bibnamefont {Kamimura}}, \bibinfo
  {author} {\bibfnamefont {M.}~\bibnamefont {Kawai}}, \bibinfo {author}
  {\bibfnamefont {G.}~\bibnamefont {Rawitscher}},\ and\ \bibinfo {author}
  {\bibfnamefont {M.}~\bibnamefont {Yahiro}},\ }\href
  {https://doi.org/10.1016/0370-1573(87)90094-9} {\bibfield  {journal}
  {\bibinfo  {journal} {Phys. Rep.}\ }\textbf {\bibinfo {volume} {154}},\
  \bibinfo {pages} {125} (\bibinfo {year} {1987})}\BibitemShut {NoStop}%
\bibitem [{\citenamefont {Gómez-Ramos}\ \emph {et~al.}(2022)\citenamefont
  {Gómez-Ramos}, \citenamefont {Gómez-Camacho},\ and\ \citenamefont
  {Moro}}]{Gom22}%
  \BibitemOpen
  \bibfield  {author} {\bibinfo {author} {\bibfnamefont {M.}~\bibnamefont
  {Gómez-Ramos}}, \bibinfo {author} {\bibfnamefont {J.}~\bibnamefont
  {Gómez-Camacho}},\ and\ \bibinfo {author} {\bibfnamefont {A.}~\bibnamefont
  {Moro}},\ }\href
  {https://doi.org/https://doi.org/10.1016/j.physletb.2022.137252} {\bibfield
  {journal} {\bibinfo  {journal} {Physics Letters B}\ }\textbf {\bibinfo
  {volume} {832}},\ \bibinfo {pages} {137252} (\bibinfo {year}
  {2022})}\BibitemShut {NoStop}%
\bibitem [{\citenamefont {Hebborn}\ and\ \citenamefont {Potel}(2023)}]{Heb23}%
  \BibitemOpen
  \bibfield  {author} {\bibinfo {author} {\bibfnamefont {C.}~\bibnamefont
  {Hebborn}}\ and\ \bibinfo {author} {\bibfnamefont {G.}~\bibnamefont
  {Potel}},\ }\href {https://doi.org/10.1103/PhysRevC.107.014607} {\bibfield
  {journal} {\bibinfo  {journal} {Phys. Rev. C}\ }\textbf {\bibinfo {volume}
  {107}},\ \bibinfo {pages} {014607} (\bibinfo {year} {2023})}\BibitemShut
  {NoStop}%
\bibitem [{\citenamefont {Sauvan}\ \emph {et~al.}(2004)\citenamefont {Sauvan},
  \citenamefont {Carstoiu}, \citenamefont {Orr}, \citenamefont {Winfield},
  \citenamefont {Freer}, \citenamefont {Ang\'elique}, \citenamefont {Catford},
  \citenamefont {Clarke}, \citenamefont {Curtis}, \citenamefont {Gr\'evy},
  \citenamefont {Le~Brun}, \citenamefont {Lewitowicz}, \citenamefont
  {Li\'egard}, \citenamefont {Marqu\'es}, \citenamefont {Mac~Cormick},
  \citenamefont {Roussel-Chomaz}, \citenamefont {Saint~Laurent},\ and\
  \citenamefont {Shawcross}}]{Sau04}%
  \BibitemOpen
  \bibfield  {author} {\bibinfo {author} {\bibfnamefont {E.}~\bibnamefont
  {Sauvan}}, \bibinfo {author} {\bibfnamefont {F.}~\bibnamefont {Carstoiu}},
  \bibinfo {author} {\bibfnamefont {N.~A.}\ \bibnamefont {Orr}}, \bibinfo
  {author} {\bibfnamefont {J.~S.}\ \bibnamefont {Winfield}}, \bibinfo {author}
  {\bibfnamefont {M.}~\bibnamefont {Freer}}, \bibinfo {author} {\bibfnamefont
  {J.~C.}\ \bibnamefont {Ang\'elique}}, \bibinfo {author} {\bibfnamefont
  {W.~N.}\ \bibnamefont {Catford}}, \bibinfo {author} {\bibfnamefont {N.~M.}\
  \bibnamefont {Clarke}}, \bibinfo {author} {\bibfnamefont {N.}~\bibnamefont
  {Curtis}}, \bibinfo {author} {\bibfnamefont {S.}~\bibnamefont {Gr\'evy}},
  \bibinfo {author} {\bibfnamefont {C.}~\bibnamefont {Le~Brun}}, \bibinfo
  {author} {\bibfnamefont {M.}~\bibnamefont {Lewitowicz}}, \bibinfo {author}
  {\bibfnamefont {E.}~\bibnamefont {Li\'egard}}, \bibinfo {author}
  {\bibfnamefont {F.~M.}\ \bibnamefont {Marqu\'es}}, \bibinfo {author}
  {\bibfnamefont {M.}~\bibnamefont {Mac~Cormick}}, \bibinfo {author}
  {\bibfnamefont {P.}~\bibnamefont {Roussel-Chomaz}}, \bibinfo {author}
  {\bibfnamefont {M.-G.}\ \bibnamefont {Saint~Laurent}},\ and\ \bibinfo
  {author} {\bibfnamefont {M.}~\bibnamefont {Shawcross}},\ }\href
  {https://doi.org/10.1103/PhysRevC.69.044603} {\bibfield  {journal} {\bibinfo
  {journal} {Phys. Rev. C}\ }\textbf {\bibinfo {volume} {69}},\ \bibinfo
  {pages} {044603} (\bibinfo {year} {2004})}\BibitemShut {NoStop}%
\bibitem [{\citenamefont {Brown}\ \emph {et~al.}(2002)\citenamefont {Brown},
  \citenamefont {Hansen}, \citenamefont {Sherrill},\ and\ \citenamefont
  {Tostevin}}]{bro02}%
  \BibitemOpen
  \bibfield  {author} {\bibinfo {author} {\bibfnamefont {B.~A.}\ \bibnamefont
  {Brown}}, \bibinfo {author} {\bibfnamefont {P.~G.}\ \bibnamefont {Hansen}},
  \bibinfo {author} {\bibfnamefont {B.~M.}\ \bibnamefont {Sherrill}},\ and\
  \bibinfo {author} {\bibfnamefont {J.~A.}\ \bibnamefont {Tostevin}},\ }\href
  {https://doi.org/10.1103/PhysRevC.65.061601} {\bibfield  {journal} {\bibinfo
  {journal} {Phys. Rev. C}\ }\textbf {\bibinfo {volume} {65}},\ \bibinfo
  {pages} {061601} (\bibinfo {year} {2002})}\BibitemShut {NoStop}%
\bibitem [{\citenamefont {Stroberg}(2016)}]{Strthesis}%
  \BibitemOpen
  \bibfield  {author} {\bibinfo {author} {\bibfnamefont {S.~R.}\ \bibnamefont
  {Stroberg}},\ }\emph {\bibinfo {title} {Single-particle structure of
  neutron-ich silicon isotopes and the breakdown of the N=28 shell closure}},\
  \href@noop {} {Ph.D. thesis},\ \bibinfo  {school} {Michigan State
  University}, \bibinfo {address} {Michigan State University, East Lansing}
  (\bibinfo {year} {2016})\BibitemShut {NoStop}%
\bibitem [{\citenamefont {Morillon}\ and\ \citenamefont
  {Romain}(2007)}]{Mor07}%
  \BibitemOpen
  \bibfield  {author} {\bibinfo {author} {\bibfnamefont {B.}~\bibnamefont
  {Morillon}}\ and\ \bibinfo {author} {\bibfnamefont {P.}~\bibnamefont
  {Romain}},\ }\href {https://doi.org/10.1103/PhysRevC.76.044601} {\bibfield
  {journal} {\bibinfo  {journal} {Phys. Rev. C}\ }\textbf {\bibinfo {volume}
  {76}},\ \bibinfo {pages} {044601} (\bibinfo {year} {2007})}\BibitemShut
  {NoStop}%
\bibitem [{\citenamefont {Brown}\ \emph {et~al.}(2018)\citenamefont {Brown},
  \citenamefont {Chadwick}, \citenamefont {Capote}, \citenamefont {Kahler},
  \citenamefont {Trkov}, \citenamefont {Herman}, \citenamefont {Sonzogni},
  \citenamefont {Danon}, \citenamefont {Carlson}, \citenamefont {Dunn},
  \citenamefont {Smith}, \citenamefont {Hale}, \citenamefont {Arbanas},
  \citenamefont {Arcilla}, \citenamefont {Bates}, \citenamefont {Beck},
  \citenamefont {Becker}, \citenamefont {Brown}, \citenamefont {Casperson},
  \citenamefont {Conlin}, \citenamefont {Cullen}, \citenamefont {Descalle},
  \citenamefont {Firestone}, \citenamefont {Gaines}, \citenamefont {Guber},
  \citenamefont {Hawari}, \citenamefont {Holmes}, \citenamefont {Johnson},
  \citenamefont {Kawano}, \citenamefont {Kiedrowski}, \citenamefont {Koning},
  \citenamefont {Kopecky}, \citenamefont {Leal}, \citenamefont {Lestone},
  \citenamefont {Lubitz}, \citenamefont {{Márquez Damián}}, \citenamefont
  {Mattoon}, \citenamefont {McCutchan}, \citenamefont {Mughabghab},
  \citenamefont {Navratil}, \citenamefont {Neudecker}, \citenamefont {Nobre},
  \citenamefont {Noguere}, \citenamefont {Paris}, \citenamefont {Pigni},
  \citenamefont {Plompen}, \citenamefont {Pritychenko}, \citenamefont
  {Pronyaev}, \citenamefont {Roubtsov}, \citenamefont {Rochman}, \citenamefont
  {Romano}, \citenamefont {Schillebeeckx}, \citenamefont {Simakov},
  \citenamefont {Sin}, \citenamefont {Sirakov}, \citenamefont {Sleaford},
  \citenamefont {Sobes}, \citenamefont {Soukhovitskii}, \citenamefont {Stetcu},
  \citenamefont {Talou}, \citenamefont {Thompson}, \citenamefont {{van der
  Marck}}, \citenamefont {Welser-Sherrill}, \citenamefont {Wiarda},
  \citenamefont {White}, \citenamefont {Wormald}, \citenamefont {Wright},
  \citenamefont {Zerkle}, \citenamefont {Žerovnik},\ and\ \citenamefont
  {Zhu}}]{Bro18}%
  \BibitemOpen
  \bibfield  {author} {\bibinfo {author} {\bibfnamefont {D.}~\bibnamefont
  {Brown}}, \bibinfo {author} {\bibfnamefont {M.}~\bibnamefont {Chadwick}},
  \bibinfo {author} {\bibfnamefont {R.}~\bibnamefont {Capote}}, \bibinfo
  {author} {\bibfnamefont {A.}~\bibnamefont {Kahler}}, \bibinfo {author}
  {\bibfnamefont {A.}~\bibnamefont {Trkov}}, \bibinfo {author} {\bibfnamefont
  {M.}~\bibnamefont {Herman}}, \bibinfo {author} {\bibfnamefont
  {A.}~\bibnamefont {Sonzogni}}, \bibinfo {author} {\bibfnamefont
  {Y.}~\bibnamefont {Danon}}, \bibinfo {author} {\bibfnamefont
  {A.}~\bibnamefont {Carlson}}, \bibinfo {author} {\bibfnamefont
  {M.}~\bibnamefont {Dunn}}, \bibinfo {author} {\bibfnamefont {D.}~\bibnamefont
  {Smith}}, \bibinfo {author} {\bibfnamefont {G.}~\bibnamefont {Hale}},
  \bibinfo {author} {\bibfnamefont {G.}~\bibnamefont {Arbanas}}, \bibinfo
  {author} {\bibfnamefont {R.}~\bibnamefont {Arcilla}}, \bibinfo {author}
  {\bibfnamefont {C.}~\bibnamefont {Bates}}, \bibinfo {author} {\bibfnamefont
  {B.}~\bibnamefont {Beck}}, \bibinfo {author} {\bibfnamefont {B.}~\bibnamefont
  {Becker}}, \bibinfo {author} {\bibfnamefont {F.}~\bibnamefont {Brown}},
  \bibinfo {author} {\bibfnamefont {R.}~\bibnamefont {Casperson}}, \bibinfo
  {author} {\bibfnamefont {J.}~\bibnamefont {Conlin}}, \bibinfo {author}
  {\bibfnamefont {D.}~\bibnamefont {Cullen}}, \bibinfo {author} {\bibfnamefont
  {M.-A.}\ \bibnamefont {Descalle}}, \bibinfo {author} {\bibfnamefont
  {R.}~\bibnamefont {Firestone}}, \bibinfo {author} {\bibfnamefont
  {T.}~\bibnamefont {Gaines}}, \bibinfo {author} {\bibfnamefont
  {K.}~\bibnamefont {Guber}}, \bibinfo {author} {\bibfnamefont
  {A.}~\bibnamefont {Hawari}}, \bibinfo {author} {\bibfnamefont
  {J.}~\bibnamefont {Holmes}}, \bibinfo {author} {\bibfnamefont
  {T.}~\bibnamefont {Johnson}}, \bibinfo {author} {\bibfnamefont
  {T.}~\bibnamefont {Kawano}}, \bibinfo {author} {\bibfnamefont
  {B.}~\bibnamefont {Kiedrowski}}, \bibinfo {author} {\bibfnamefont
  {A.}~\bibnamefont {Koning}}, \bibinfo {author} {\bibfnamefont
  {S.}~\bibnamefont {Kopecky}}, \bibinfo {author} {\bibfnamefont
  {L.}~\bibnamefont {Leal}}, \bibinfo {author} {\bibfnamefont {J.}~\bibnamefont
  {Lestone}}, \bibinfo {author} {\bibfnamefont {C.}~\bibnamefont {Lubitz}},
  \bibinfo {author} {\bibfnamefont {J.}~\bibnamefont {{Márquez Damián}}},
  \bibinfo {author} {\bibfnamefont {C.}~\bibnamefont {Mattoon}}, \bibinfo
  {author} {\bibfnamefont {E.}~\bibnamefont {McCutchan}}, \bibinfo {author}
  {\bibfnamefont {S.}~\bibnamefont {Mughabghab}}, \bibinfo {author}
  {\bibfnamefont {P.}~\bibnamefont {Navratil}}, \bibinfo {author}
  {\bibfnamefont {D.}~\bibnamefont {Neudecker}}, \bibinfo {author}
  {\bibfnamefont {G.}~\bibnamefont {Nobre}}, \bibinfo {author} {\bibfnamefont
  {G.}~\bibnamefont {Noguere}}, \bibinfo {author} {\bibfnamefont
  {M.}~\bibnamefont {Paris}}, \bibinfo {author} {\bibfnamefont
  {M.}~\bibnamefont {Pigni}}, \bibinfo {author} {\bibfnamefont
  {A.}~\bibnamefont {Plompen}}, \bibinfo {author} {\bibfnamefont
  {B.}~\bibnamefont {Pritychenko}}, \bibinfo {author} {\bibfnamefont
  {V.}~\bibnamefont {Pronyaev}}, \bibinfo {author} {\bibfnamefont
  {D.}~\bibnamefont {Roubtsov}}, \bibinfo {author} {\bibfnamefont
  {D.}~\bibnamefont {Rochman}}, \bibinfo {author} {\bibfnamefont
  {P.}~\bibnamefont {Romano}}, \bibinfo {author} {\bibfnamefont
  {P.}~\bibnamefont {Schillebeeckx}}, \bibinfo {author} {\bibfnamefont
  {S.}~\bibnamefont {Simakov}}, \bibinfo {author} {\bibfnamefont
  {M.}~\bibnamefont {Sin}}, \bibinfo {author} {\bibfnamefont {I.}~\bibnamefont
  {Sirakov}}, \bibinfo {author} {\bibfnamefont {B.}~\bibnamefont {Sleaford}},
  \bibinfo {author} {\bibfnamefont {V.}~\bibnamefont {Sobes}}, \bibinfo
  {author} {\bibfnamefont {E.}~\bibnamefont {Soukhovitskii}}, \bibinfo {author}
  {\bibfnamefont {I.}~\bibnamefont {Stetcu}}, \bibinfo {author} {\bibfnamefont
  {P.}~\bibnamefont {Talou}}, \bibinfo {author} {\bibfnamefont
  {I.}~\bibnamefont {Thompson}}, \bibinfo {author} {\bibfnamefont
  {S.}~\bibnamefont {{van der Marck}}}, \bibinfo {author} {\bibfnamefont
  {L.}~\bibnamefont {Welser-Sherrill}}, \bibinfo {author} {\bibfnamefont
  {D.}~\bibnamefont {Wiarda}}, \bibinfo {author} {\bibfnamefont
  {M.}~\bibnamefont {White}}, \bibinfo {author} {\bibfnamefont
  {J.}~\bibnamefont {Wormald}}, \bibinfo {author} {\bibfnamefont
  {R.}~\bibnamefont {Wright}}, \bibinfo {author} {\bibfnamefont
  {M.}~\bibnamefont {Zerkle}}, \bibinfo {author} {\bibfnamefont
  {G.}~\bibnamefont {Žerovnik}},\ and\ \bibinfo {author} {\bibfnamefont
  {Y.}~\bibnamefont {Zhu}},\ }\href
  {https://doi.org/https://doi.org/10.1016/j.nds.2018.02.001} {\bibfield
  {journal} {\bibinfo  {journal} {Nuclear Data Sheets}\ }\textbf {\bibinfo
  {volume} {148}},\ \bibinfo {pages} {1} (\bibinfo {year} {2018})},\ \bibinfo
  {note} {special Issue on Nuclear Reaction Data}\BibitemShut {NoStop}%
\bibitem [{\citenamefont {Blank}\ \emph {et~al.}(2018)\citenamefont {Blank},
  \citenamefont {Canchel}, \citenamefont {Seis},\ and\ \citenamefont
  {Delahaye}}]{Bla18}%
  \BibitemOpen
  \bibfield  {author} {\bibinfo {author} {\bibfnamefont {B.}~\bibnamefont
  {Blank}}, \bibinfo {author} {\bibfnamefont {G.}~\bibnamefont {Canchel}},
  \bibinfo {author} {\bibfnamefont {F.}~\bibnamefont {Seis}},\ and\ \bibinfo
  {author} {\bibfnamefont {P.}~\bibnamefont {Delahaye}},\ }\href
  {https://doi.org/https://doi.org/10.1016/j.nimb.2017.12.003} {\bibfield
  {journal} {\bibinfo  {journal} {Nuclear Instruments and Methods in Physics
  Research Section B: Beam Interactions with Materials and Atoms}\ }\textbf
  {\bibinfo {volume} {416}},\ \bibinfo {pages} {41} (\bibinfo {year}
  {2018})}\BibitemShut {NoStop}%
\bibitem [{\citenamefont {Tarasov}\ and\ \citenamefont {Bazin}(2008)}]{Tar08}%
  \BibitemOpen
  \bibfield  {author} {\bibinfo {author} {\bibfnamefont {O.}~\bibnamefont
  {Tarasov}}\ and\ \bibinfo {author} {\bibfnamefont {D.}~\bibnamefont
  {Bazin}},\ }\href
  {https://doi.org/https://doi.org/10.1016/j.nimb.2008.05.110} {\bibfield
  {journal} {\bibinfo  {journal} {Nuclear Instruments and Methods in Physics
  Research Section B: Beam Interactions with Materials and Atoms}\ }\textbf
  {\bibinfo {volume} {266}},\ \bibinfo {pages} {4657} (\bibinfo {year}
  {2008})},\ \bibinfo {note} {proceedings of the XVth International Conference
  on Electromagnetic Isotope Separators and Techniques Related to their
  Applications}\BibitemShut {NoStop}%
\bibitem [{\citenamefont {Gavron}(1980)}]{Gav80}%
  \BibitemOpen
  \bibfield  {author} {\bibinfo {author} {\bibfnamefont {A.}~\bibnamefont
  {Gavron}},\ }\href {https://doi.org/10.1103/PhysRevC.21.230} {\bibfield
  {journal} {\bibinfo  {journal} {Phys. Rev. C}\ }\textbf {\bibinfo {volume}
  {21}},\ \bibinfo {pages} {230} (\bibinfo {year} {1980})}\BibitemShut
  {NoStop}%
\bibitem [{\citenamefont {Charity}(2010)}]{Cha10}%
  \BibitemOpen
  \bibfield  {author} {\bibinfo {author} {\bibfnamefont {R.~J.}\ \bibnamefont
  {Charity}},\ }\href {https://doi.org/10.1103/PhysRevC.82.014610} {\bibfield
  {journal} {\bibinfo  {journal} {Phys. Rev. C}\ }\textbf {\bibinfo {volume}
  {82}},\ \bibinfo {pages} {014610} (\bibinfo {year} {2010})}\BibitemShut
  {NoStop}%
\bibitem [{\citenamefont {Mancusi}\ \emph {et~al.}(2010)\citenamefont
  {Mancusi}, \citenamefont {Charity},\ and\ \citenamefont {Cugnon}}]{Man10}%
  \BibitemOpen
  \bibfield  {author} {\bibinfo {author} {\bibfnamefont {D.}~\bibnamefont
  {Mancusi}}, \bibinfo {author} {\bibfnamefont {R.~J.}\ \bibnamefont
  {Charity}},\ and\ \bibinfo {author} {\bibfnamefont {J.}~\bibnamefont
  {Cugnon}},\ }\href {https://doi.org/10.1103/PhysRevC.82.044610} {\bibfield
  {journal} {\bibinfo  {journal} {Phys. Rev. C}\ }\textbf {\bibinfo {volume}
  {82}},\ \bibinfo {pages} {044610} (\bibinfo {year} {2010})}\BibitemShut
  {NoStop}%
\bibitem [{nud()}]{nudat}%
  \BibitemOpen
  \href {{https://www.nndc.bnl.gov/nudat/}} {\bibinfo {title} {Nudat database,
  national nuclear data center}}\BibitemShut {NoStop}%
\bibitem [{\citenamefont {Jensen}\ \emph {et~al.}(2011)\citenamefont {Jensen},
  \citenamefont {Hagen}, \citenamefont {Hjorth-Jensen}, \citenamefont {Brown},\
  and\ \citenamefont {Gade}}]{Jen11}%
  \BibitemOpen
  \bibfield  {author} {\bibinfo {author} {\bibfnamefont {O.}~\bibnamefont
  {Jensen}}, \bibinfo {author} {\bibfnamefont {G.}~\bibnamefont {Hagen}},
  \bibinfo {author} {\bibfnamefont {M.}~\bibnamefont {Hjorth-Jensen}}, \bibinfo
  {author} {\bibfnamefont {B.~A.}\ \bibnamefont {Brown}},\ and\ \bibinfo
  {author} {\bibfnamefont {A.}~\bibnamefont {Gade}},\ }\href
  {https://doi.org/10.1103/PhysRevLett.107.032501} {\bibfield  {journal}
  {\bibinfo  {journal} {Phys. Rev. Lett.}\ }\textbf {\bibinfo {volume} {107}},\
  \bibinfo {pages} {032501} (\bibinfo {year} {2011})}\BibitemShut {NoStop}%
\bibitem [{\citenamefont {Idini}\ \emph {et~al.}(2019)\citenamefont {Idini},
  \citenamefont {Barbieri},\ and\ \citenamefont {Navr\'atil}}]{Idi19}%
  \BibitemOpen
  \bibfield  {author} {\bibinfo {author} {\bibfnamefont {A.}~\bibnamefont
  {Idini}}, \bibinfo {author} {\bibfnamefont {C.}~\bibnamefont {Barbieri}},\
  and\ \bibinfo {author} {\bibfnamefont {P.}~\bibnamefont {Navr\'atil}},\
  }\href {https://doi.org/10.1103/PhysRevLett.123.092501} {\bibfield  {journal}
  {\bibinfo  {journal} {Phys. Rev. Lett.}\ }\textbf {\bibinfo {volume} {123}},\
  \bibinfo {pages} {092501} (\bibinfo {year} {2019})}\BibitemShut {NoStop}%
\bibitem [{\citenamefont {Ichimura}\ \emph {et~al.}(1985)\citenamefont
  {Ichimura}, \citenamefont {Austern},\ and\ \citenamefont {Vincent}}]{IAV85}%
  \BibitemOpen
  \bibfield  {author} {\bibinfo {author} {\bibfnamefont {M.}~\bibnamefont
  {Ichimura}}, \bibinfo {author} {\bibfnamefont {N.}~\bibnamefont {Austern}},\
  and\ \bibinfo {author} {\bibfnamefont {C.~M.}\ \bibnamefont {Vincent}},\
  }\href {https://doi.org/10.1103/PhysRevC.32.431} {\bibfield  {journal}
  {\bibinfo  {journal} {Phys. Rev. C}\ }\textbf {\bibinfo {volume} {32}},\
  \bibinfo {pages} {431} (\bibinfo {year} {1985})}\BibitemShut {NoStop}%
\bibitem [{\citenamefont {Lei}\ and\ \citenamefont {{Moro}}(2015)}]{Jin15}%
  \BibitemOpen
  \bibfield  {author} {\bibinfo {author} {\bibfnamefont {J.}~\bibnamefont
  {Lei}}\ and\ \bibinfo {author} {\bibfnamefont {A.~M.}\ \bibnamefont
  {{Moro}}},\ }\href {https://doi.org/10.1103/PhysRevC.92.044616} {\bibfield
  {journal} {\bibinfo  {journal} {Phys. Rev. C}\ }\textbf {\bibinfo {volume}
  {92}},\ \bibinfo {pages} {044616} (\bibinfo {year} {2015})}\BibitemShut
  {NoStop}%
\bibitem [{\citenamefont {Lei}\ and\ \citenamefont {Moro}(2015)}]{Jin15b}%
  \BibitemOpen
  \bibfield  {author} {\bibinfo {author} {\bibfnamefont {J.}~\bibnamefont
  {Lei}}\ and\ \bibinfo {author} {\bibfnamefont {A.~M.}\ \bibnamefont {Moro}},\
  }\href {https://doi.org/10.1103/PhysRevC.92.061602} {\bibfield  {journal}
  {\bibinfo  {journal} {Phys. Rev. C}\ }\textbf {\bibinfo {volume} {92}},\
  \bibinfo {pages} {061602} (\bibinfo {year} {2015})}\BibitemShut {NoStop}%
\bibitem [{\citenamefont {Potel}\ \emph {et~al.}(2015)\citenamefont {Potel},
  \citenamefont {Nunes},\ and\ \citenamefont {Thompson}}]{Pot15}%
  \BibitemOpen
  \bibfield  {author} {\bibinfo {author} {\bibfnamefont {G.}~\bibnamefont
  {Potel}}, \bibinfo {author} {\bibfnamefont {F.~M.}\ \bibnamefont {Nunes}},\
  and\ \bibinfo {author} {\bibfnamefont {I.~J.}\ \bibnamefont {Thompson}},\
  }\href {https://doi.org/10.1103/PhysRevC.92.034611} {\bibfield  {journal}
  {\bibinfo  {journal} {Phys. Rev. C}\ }\textbf {\bibinfo {volume} {92}},\
  \bibinfo {pages} {034611} (\bibinfo {year} {2015})}\BibitemShut {NoStop}%
\bibitem [{\citenamefont {Carlson}\ \emph {et~al.}(2016)\citenamefont
  {Carlson}, \citenamefont {Capote},\ and\ \citenamefont {Sin}}]{Car16}%
  \BibitemOpen
  \bibfield  {author} {\bibinfo {author} {\bibfnamefont {B.~V.}\ \bibnamefont
  {Carlson}}, \bibinfo {author} {\bibfnamefont {R.}~\bibnamefont {Capote}},\
  and\ \bibinfo {author} {\bibfnamefont {M.}~\bibnamefont {Sin}},\ }\href
  {https://doi.org/10.1007/s00601-016-1054-8} {\bibfield  {journal} {\bibinfo
  {journal} {Few-Body Syst.}\ }\textbf {\bibinfo {volume} {57}},\ \bibinfo
  {pages} {307} (\bibinfo {year} {2016})}\BibitemShut {NoStop}%
\end{thebibliography}%

\end{document}


\maketitle

\section{Derivation of the approximate expression of the product of S-matrices for different core-valence coordinates}

 The expression of the proton removal probability will require integration over two radial variables, $\vecru, \vecrd$. The integrand would involve the product of two S matrices $S^j_{VT}(b_{VT}(\vecru)) (S^j_{VT}(b_{VT}(\vecrd)))^*$, which are evaluated at different impact parameters, so unitarity (see Eq.~(3) of the manuscript) cannot be applied. This problem with unitarity can be avoided approximating the two impact parameters in the previous expressions by an average impact parameter defined as $b_{VT}= \sqrt{(b + \alpha x)^2 + (\alpha y)^2}$
where  $\alpha = {A-1 \over A}$, $x= {x_1 + x_2 \over 2}$ and $y = {\sqrt{\dfrac{y_1 ^2 + y_2^2}{2}}}$. 

Here, we will indicate how this derivation is obtained. The value of $S^j_{VT}(b_{VT}(\vecru))$ can be written, to first order in $ b_{VT}(\vecru)-b_{VT} $ as
\begin{equation}
S^j_{VT}(b_{VT}(\vecru)) \simeq S^j_{VT}(b_{VT}) e^{2 (\alpha^j_{VT} + i \beta^j_{VT})  (b_{VT}(\vecru)-b_{VT})} ,
\end{equation}
where $2 (\alpha^j_{VT}+ i \beta^j_{VT}) $ is the logarithmic derivative of $S^j_{VT}(b_{VT})$ with respect to the impact parameter of the valence particle  with respect to the target $b_{VT}$. Note that if we express $S^j_{VT}(b_{VT}) = \exp(2i \delta^j)$ in terms of the phaseshift then $\alpha^j_{VT}$ is the derivative of the imaginary part of the phaseshift, while then $\beta^j_{VT}$ is the derivative of the real part of the phaseshift,
Similarly, 
\begin{equation}
(S^j_{VT}(b_{VT}(\vecrd)))^* \simeq (S^j_{VT}(b_{VT}))^* e^{ 2 (\alpha^j_{VT} - i \beta^j_{VT})  (b_{VT}(\vecrd)-b_{VT})} .
\end{equation}
Then, the product of the two S-matrices is given by
\begin{equation}
S^j_{VT}(b_{VT}(\vecru)) (S^j_{VT}(b_{VT}(\vecrd)))^* \simeq |S^j_{VT}(b_{VT})|^2 e^{2 \alpha^j_{VT} (b_{VT}(\vecru)+ b_{VT}(\vecrd)-2 b_{VT})} e^{ 2 i \beta^j_{VT}  (b_{VT}(\vecru)-b_{VT}(\vecrd))}
\end{equation}
Note that a judicious choice of $b_{VT}$ can minimize the effect of the first exponential. In fact, a Taylor expansion of  $b_{VT}(\vecru)$, $b_{VT}(\vecrd)$ and  $b_{VT}$ in terms of $x_1, y_1, x_2, y_2, x, y$, which are considered small compared to the impact parameter $b$, yields
\begin{equation}
  (b_{VT}(\vecru)+ b_{VT}(\vecrd)-2 b_{VT}) \simeq \alpha (x_1 + x_2 - 2 x) + {\alpha^2 \over 2 b} (y_1^2 + y_2^2 - 2 y^2)   ,
\end{equation}
so, this justifies the choice $x= (x_1+x_2)/2$ and $y = \sqrt{(y_1^2 + y_2^2)/2}$,  which makes $  (b_{VT}(\vecru)+ b_{VT}(\vecrd)-2 b_{VT}) \simeq 0$, and hence, for arbitrary $ \alpha^j_{VT}  $,  $e^{2 \alpha^j_{VT} (b_{VT}(\vecru)+ b_{VT}(\vecrd)-2 b_{VT})} \simeq 1$.

The second exponential introduces a phase that is independent of the choice of $b_{VT}$.  $\beta^j_{VT}$ is the derivative of the phaseshift with respect to the impact parameter, and the phaseshift is the integral of the real potential with respect to the time. As the radial derivative of the potential is a force, and the integral on time of the force is a momentum, $\beta^j_{VT}$ can be seen as a transverse momentum given by the target to the valence particle. We can argue that, if the energy of the collision is sufficiently large, then the collision time is small, and so the factor $\beta^j_{VT}$ becomes small for large collision energies.
This cannot be said about the factor $\alpha^j_{VT}$, which indicates the dependence of the imaginary phaseshift with the impact parameter. This dependence will be relevant at all collision energies, as  we expect that larger impact parameters will lead to valence survival, and smaller ones to strong absorption. However, as previously shown, a judicious choice of $x, y$ cancels the dependence on $\alpha^j_{VT}$.

With these considerations, we can approximate
\begin{equation}
S^j_{VT}(b_{VT}(\vecru)) (S^j_{VT}(b_{VT}(\vecrd)))^* \simeq |S^j_{VT}(b_{VT})|^2 ,
\end{equation}
provided that $b_{VT} = \sqrt{ (b + \alpha x)^2 + (\alpha y)^2}$, taking $x= (x_1+x_2)/2$ and $y = \sqrt{(y_1^2 + y_2^2)/2}$.

\section{Computation of $\rho^{(2)\rm{eff}}$}
Let us start with the expression for $\rho^{(2)\rm{eff}}(x,y)$ given in the main document: 
\begin{align}
\label{eq:efstr}\rho^{(2)\rm{eff}}(x,y) &= \int \! d^3 \vecru \int \! d^3 \vecrd \; \delta(x- {x_1 + x_2 \over 2})  \delta(y- \sqrt{\dfrac{y_1^2 + y_2^2}{2}}) \nonumber \\ &\times \langle \vecrd |\rho_f | \vecru \rangle \phi_g^*(\vecrd) \phi_g(\vecru)\\
 \langle \vecrd |\rho_f | \vecru \rangle &=  \int d^3 \veck \; \left( \psi^{(-)}(\vec k, \vecrd) \right)^*  \psi^{(-)}(\vec k, \vecru)  .
 \label{closure}
\end{align}

In order to constrain the integration over $\mathbf{k}$ we introduce a factor $\exp(- k^4 a^4)$ where $a$ is taken as small as possible.

A multipole expansion can be carried out for $\langle \vecrd |\rho_f | \vecru \rangle$, and the integration for the angles $\hat k$ can be analytically performed, using
\begin{equation}
  \psi^{(-)}_{0}(\vec k, \vecr) = \sum_{\ell m} f_\ell(k,r) Y^*_{\ell m}(\vecr)  Y_{\ell m}(\vec k)  
\end{equation}
so
\begin{align}
\langle \vecru |\rho_f(a) | \vecrd \rangle
     &= \sum_{\ell} {2 \ell + 1 \over 4 \pi} P_\ell(\hat r_2 \cdot \hat r_1)  G_\ell(a; r_2, r_1)  \\
    G_\ell(a; r_2, r_1) & =
    \int_0^\infty k^2 dk  \; \exp(-k^4 a^4)   f_\ell^*(k, r_2)  f_\ell(k, r_1)  .
\end{align}

We have derived this expression neglecting spin-dependent forces. However, these may be included, so that the interaction, as well as the scattering wavefunction, depend on the channels spin $s$ as well as on the orbital angular momentum $\ell$ and the total angular momentum $j$ of the projectile. We get for this general case:
\begin{eqnarray}\langle \vecru |\rho_f(a) | \vecrd \rangle
     &=& \sum_{j \ell s} {2 j + 1 \over (2s+1) 4 \pi} P_\ell(\hat r_2 \cdot \hat r_1)  G_j^{(\ell s)}(a; r_2, r_1)  \\
    G_j^{(\ell s)}(a; r_2, r_1) & =&
    \int_0^\infty k^2 dk  \; \exp(-k^4 a^4)   f_j^{(\ell s)*}(k, r_2)  f_j^{(\ell s)}(k, r_1).  
\end{eqnarray}

A similar multipole expansion can be performed for $\phi_g^*(\vecrd) \phi_g(\vecru)$, leading to

\begin{eqnarray}
\rho^{(2)\rm{eff}}(x,y) &=&    \sum_{j \ell \ell_g s} {2 j + 1 \over (2s+1)(4\pi)^2}
\int d \vecru \int d \vecrd \delta(\bar{x})\delta(\bar{y})    \; P_{\ell}( \hat r_1 \cdot \hat r_2) P_{\ell_g}( \hat r_1 \cdot \hat r_2)   \\
&\times& G_j^{(\ell s)}(a; r_1, r_2) \varphi_g^{\ell_g s}(r_1)\varphi_g^{\ell_g s *}(r_2),
\end{eqnarray}
where $\delta(\bar{x})=\delta(x- {x_1 + x_2 \over 2})$ and $\delta(\bar{y})=\delta(y- \sqrt{\dfrac{y_1^2 + y_2^2}{2}})$. It is convenient to describe explicitly this sixfold integral in cartesian coordinates, which we label  ranging all by $d^6X$ in the interval $(-\infty, \infty)$.  The integrals in $y_i$ and $z_i$ can be restricted to the positive range $(0, \infty)$, taking into account the effect that the sign change has on the polynomials. Thus, we get
\begin{align}
\rho^{(2)\rm{eff}}(x,y) &=  \int d^6 X \sum\limits_{j \ell \ell_g s} \delta(\bar{x})  \delta(\bar{y})  G_j^{(\ell s)}(a; r_1, r_2) \varphi_g^{\ell_g s}(r_1)\varphi_g^{\ell_g s *}(r_2)
Q^{(\ell \ell_g)}_j(X) \\
Q^{(\ell \ell_g)}_j(X) &= 4 {2 j + 1 \over(2s+1) (4\pi)^2} \left\{ P_{\ell}(a)  P_{\ell_g}(a) + P_{\ell}(b)  P_{\ell_g}(b) +P_{\ell}(c)  P_{\ell_g}(c) + P_{\ell}(d)  P_{\ell_g}(d) \right\} ,
\end{align}
with 
\begin{align}
a&={x_1 x_2 + y_1 y_2 + z_1 z_2 \over r_1 r_2} \\ 
 b&={x_1 x_2 - y_1 y_2 + z_1 z_2 \over r_1 r_2} \\
  c&={x_1 x_2 + y_1 y_2 - z_1 z_2 \over r_1 r_2} \\
   d&={x_1 x_2 - y_1 y_2 - z_1 z_2 \over r_1 r_2} 
\end{align}
It is convenient to define new variables to perform the integration, noticing that the profile factors $(1-|S_{VT}|^2) |S_{CT}|^2$, as indicated in the main text, depend only on $x=(x_1+x_2)/2$ and $y$ fulfilling 
$2 (x^2+y^2) = u_1^2 + u_2^2$. So, instead of the variables $X = (x_1, x_2, y_1, y_2, z_1, z_2) $ we take $x=(x_1+x_2)/2$, which is defined in the interval $(-\infty, \infty)$;  $y$ fulfilling 
$2 (x^2+y^2) = u_1^2 + u_2^2$, which is in the interval $(0, \infty)$;
$w=x_1-x_2$, which is in the interval  $(-\infty, \infty)$; $\theta$ given by the conditions $y_1 = \sqrt{2}y \sin \theta$ and  $y_2 = \sqrt{2}y \cos \theta$, which is in the interval $(0, \pi/2)$ and $z_1$ and $z_2$ are left unaffected. The Jacobian of these transformations is $2 y$. Thus, one gets

\begin{align}
\rho^{(2)\rm{eff}}(x,y) &=  2y \int_{-\infty}^{\infty} dw \; 
\int_{0}^{\pi/2} d\theta 
\int_0^\infty d z_1 \int_0^\infty dz_2 \; G_j^{(\ell s)}(a; r_1, r_2) \varphi_g^{\ell_g s}(r_1)\varphi_g^{\ell_g s *}(r_2)  Q^{(\ell \ell_g)}_j(X).
\end{align}
\normalsize

The corresponding cross section can be obtained integrating the scattering probability for all impact parameters. 
\begin{align}
    \sigma &= \int_0^{\infty} 2 \pi b\,  db\, P_P(b) \simeq 2 \int_{-\infty}^\infty dx \int_{0}^\infty dy \; \rho^{(2)eff}(x,y)  \int_0^{\infty} 2 \pi b \, db \, (1- |S_{VT}|^2) |S_{CT}|^2 
\end{align}

\section{Subtraction of elastic-compound nucleus cross sections from reaction cross sections  }

\begin{figure}[tb]
\begin{center}
 {\centering \resizebox*{\columnwidth}{!}{\includegraphics{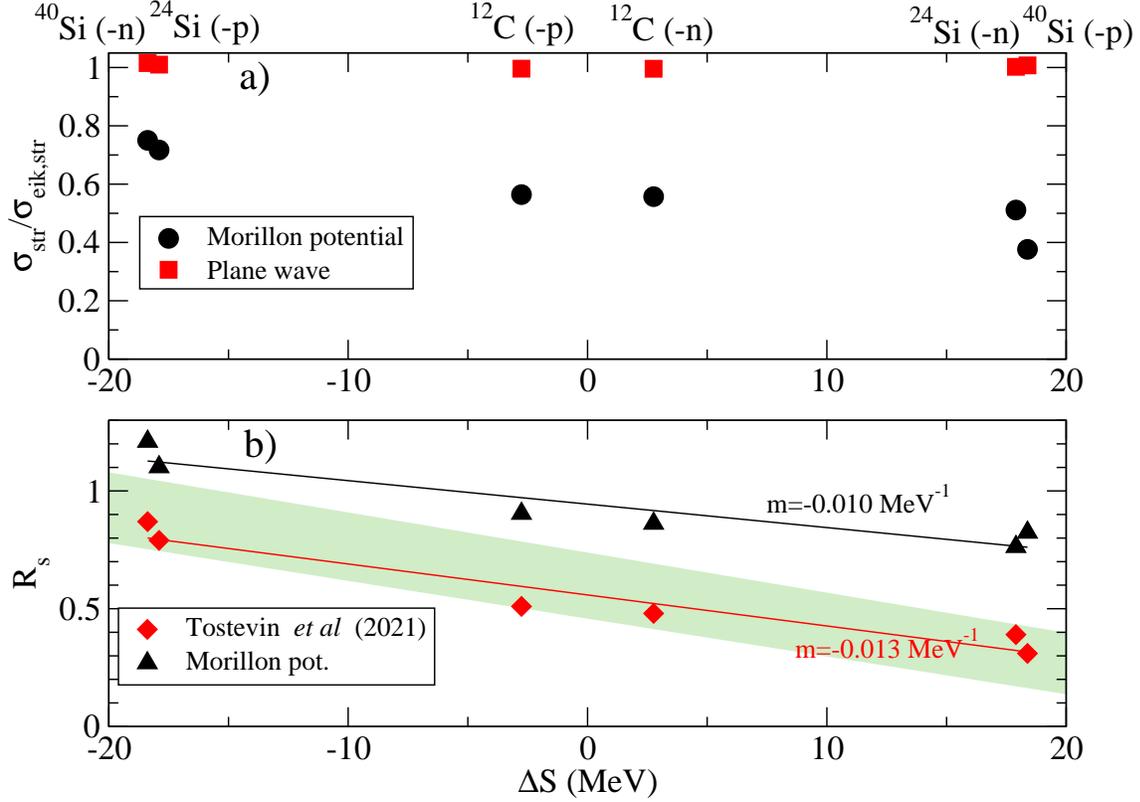}} \par}
\caption{\label{fig:gademor} a) Ratio between computed stripping cross sections and standard eikonal calculations as a function of the difference in proton-neutron separation energy $\Delta S$. Calculations are shown without the $N+C$ potential (red squares) and with the $N+C$ potential of Morillon (black circles). b) Standard (red diamonds) and modified ``quenching factors'' $R_s$ as a function of $\Delta S$, considering for the $N+C$ system the potential by Morillon (black triangles).}
\end{center}
\end{figure}

In Figs.~\ref{fig:densmor} and \ref{fig:gademor} results are presented in which the optical potential has been taken as the imaginary part of the potential by Morillon \textit{et al}, without readjusting to remove the contribution of compound nucleus formation leading to the core-nucleon channel.

\begin{figure}[tb]
\begin{center}
 {\centering \resizebox*{\columnwidth}{!}{\includegraphics{densmorillon.eps}} \par}
\caption{\label{fig:densmor} One-dimensional effective density for the valence neutron (left) and valence proton (right) in the $1f_{7/2}$ and $1d_{5/2}$ orbitals, respectively, for $^{40}$Si. The red line corresponds to the effective density without core destruction, and the orange one to the eikonal calculation. The black curve corresponds to the calculation with core destruction, modelled with the dispersive potential by Morillon.}
\end{center}
\end{figure}

As mentioned in the main text, the results for a large (positive) $\Delta S$ did not need modification to remove the elastic contribution of the compound nucleus formation. As such, the results for $^{24}$Si$(-n)$ and $^{40}$Si$(-p)$ are the same as those from the main text. Meanwhile, for negative $\Delta S$ the difference is significant, resulting in ``quenching factors'' greater than unity. We note that the Morillon potential (as other global potentials) predicts a finite reaction cross section even at zero relative energy between nucleon and core even for the weakly-bound cases, where no open channels exist at this energy, apart from the elastic channel (and the bound state, which can only be reached by radiative capture, a process not considered in global optical potentials). As mentioned in the main text, this reaction cross section corresponds to the formation of the compound nucleus, which, for the weakly-bound cases, can only decay in the core-nucleon channel and, as such, does not remove flux from this channel, so its inclusion results in an artificially high core destruction (and artificially high $R_s$). In order to remove this contribution, the method with compound nucleus calculations described in the main text was used.

\section{Scaling factors for the surface term of the potential}

In Table~\ref{tab:factors} the scaling factors used to rescale the imaginary part of the Morillon potential are presented as a function of the nucleon-core relative energy. These factors were fitted  with a powered Woods-Saxon type function for interpolation. When the reaction cross section obtained from compound-nucleus calculations was too small, the imaginary volume term had to be nullified. These cases are denoted by an asterisk. It should be noted that for the deeply-bound nucleons ($p+^{39}$Al and $n+^{23}$Si) no modification was required, so all factors were 1. It is also notable that the calculations with GEMINI give systematically larger factors than those with PACE.

\begin{table}[!ht]
    \centering
    \begin{tabular}{|c|cc|cc|cc|cc|}
    \hline
        ~&\multicolumn{2}{c|}{$n+^{39}$Si} & \multicolumn{2}{c|}{$p+^{23}$Al} & \multicolumn{2}{c|}{$p+^{11}$B} &  \multicolumn{2}{c|}{$n+^{11}$C} \\ \hline
        Energy (MeV) & PACE & GEMINI & PACE & GEMINI & PACE & GEMINI & PACE & GEMINI \\ \hline
        5 & 0.000* & 0.034* & 0.000* & 0.030 & 0.000* & 0.001 & 0.092 & 0.138 \\ 
        10 & 0.003* & 0.160 & 0.000* & 0.054 & 0.550 & 0.209 & 0.757 & 0.504 \\
        15 & 0.637 & 0.854 & 0.022* & 1.000 & 0.347 & 0.468 & 0.672 & 0.660 \\ 
        20 & 0.920 & 0.991 & 0.278 & 1.000 & 0.814 & 0.677 & 0.875 & 0.801 \\ 
        25 & 1.000 & 1.000 & 0.685 & 1.000 & 0.914 & 0.865 & 0.891 & 0.918 \\ 
        30 & 1.000 & 1.000 & 0.912 & 1.000 & 1.000 & 1.000 & 0.963 & 0.973 \\ 
        35 & 1.000 & 1.000 & 1.000 & 1.000 & 1.000 & 1.000 & 1.000 & 1.000 \\ 
        40 & 1.000 & 1.000 & 1.000 & 1.000 & 1.000 & 1.000 & 1.000 & 1.000 \\ 
        45 & 1.000 & 1.000 & 1.000 & 1.000 & 1.000 & 1.000 & 1.000 & 1.000 \\ 
        50 & 1.000 & 1.000 & 1.000 & 1.000 & 1.000 & 1.000 & 1.000 & 1.000 \\ 
        55 & 1.000 & 1.000 & 1.000 & 1.000 & 1.000 & 1.000 & 1.000 & 1.000 \\ \hline
    \end{tabular}
    \caption{\label{tab:factors}Scaling factors for the surface imaginary part of the optical potential as a function of the core-nucleon relative energy. For the cases with $*$ the volume part was set to 0.}
\end{table}